\documentclass[twoside,11pt]{article}
\usepackage{hyperref}
\usepackage{jair, theapa, rawfonts}

\usepackage{listings}
\usepackage{amsmath,amsfonts,amssymb}
\usepackage{tikz}
\usepackage{subfig}
\usepackage{array}
\usepackage{pifont}
\usepackage[nolist,nohyperlinks]{acronym}
\usepackage{multirow}
\usepackage{pgfplotstable}
\usepackage{longtable}
\usepackage{xspace}
\usepackage{xcolor}
\usepackage{url}
\usepackage[toc,page]{appendix}
\usepackage{enumitem}	
\setlist{nolistsep}		

\newtheorem{definition}{Definition}

\usetikzlibrary{fit, spy, positioning, calc, arrows, arrows.meta,
  decorations.pathreplacing, decorations.markings, decorations.text, shapes, tikzmark}

\newcolumntype{M}[1]{>{\centering\arraybackslash}m{#1}}

\newcommand{\code}[1]{\texttt{#1}}%

\definecolor{keywordCommentColor}{rgb}{0.090000, 0.55, 0.20}
\definecolor{stringColor}{rgb}{0.558215, 0.000000, 0.135316}
\definecolor{typeColor}{rgb}{0.6, 0.000000, 0.3}
\definecolor{localColor}{rgb}{0.6, 0.000000, 0.3}
\definecolor{ndkeywordColor}{rgb}{0.0, 0.558215, 0.558215} 
\definecolor{commentsColor}{rgb}{0.0, 0.558215, 0.558215} 
\definecolor{keywordColor}{rgb}{0.000000, 0.000000, 0.635294}
\definecolor{newgray}{rgb}{0.3, 0.3, 0.3}
\definecolor{agreen}{rgb}{0.0, 0.36, 0.15}
\definecolor{rgreen}{rgb}{0.13, 0.26, 0.12}

\usepackage{refcount}
\newcounter{nalg}

\renewcommand{\thenalg}{\arabic{nalg}} 
\DeclareCaptionLabelFormat{algocaption}{Algorithm \thenalg}

\lstnewenvironment{algorithm}[1][] 
                  {
    \setcounter{lstlisting}{\value{nalg}}
    \refstepcounter{nalg}
    \captionsetup{labelformat=algocaption, labelsep=colon}
    \lstset{ 
       mathescape=true,
       escapeinside=@@,       
       comment = [l]{//},
       commentstyle=\itshape\ttfamily\textcolor{commentsColor},
       frame=tb,
       numbers=left, 
       numberstyle=\scriptsize,
       basicstyle=\footnotesize, 
       keywords={,input, min, output, return, solve, datatype, function, in,
       if, else, foreach, not, while, endwhile, begin, end, relax, for, spawn,
       psolve, valid, combine, },
       keywordstyle=\color{keywordColor}\bfseries\ttfamily\em,            
       morekeywords = [2]{solve_lns, term_cond, solve_decomp_lns},
       keywordstyle = [2]\color{keywordCommentColor}\bfseries, 
       xleftmargin=.04\textwidth,
       #1 
    }
}
{}

\newcommand{\basicCodeStyle}{\ttfamily\scriptsize\color{newgray}}
\lstdefinestyle{mipsstyle}{%
  comment = [l]{\#},
  escapeinside={!!},
  frame=tb,
  keywordstyle=\color{keywordColor}\bfseries\ttfamily\em,
  identifierstyle=\color{black}\ttfamily,
  commentstyle=\itshape\ttfamily\textcolor{commentsColor},
  stringstyle=\color{Mahogany}\ttfamily,
  numberstyle=\scriptsize,
  basicstyle= \basicCodeStyle,
  keywords={,xori, andi, b, lbu, lui, addu, lw, sw, add, addi,
  addiu, jr, jal, nop, move, slti, beqz, blez, mul, },
  morekeywords = [2]{\$zero,\$sp,\$ra,\$a0,\$a1,\$a2,\$a3,\$v0,\$v1,
  \$t0,\$t1,\$t2,\$t3,\$t4,\$t5,\$t6,\$t7,\$t8,
  \$t9,\$s1,\$s2,\$s3,\$s4,\$s5, \$fp,\$gp},
  keywordstyle = [2]\color{keywordCommentColor}\bfseries, 
  numbers=left,
  xleftmargin=.08\textwidth,
}

\lstdefinestyle{cstyle}{%
  language=c,
  basicstyle= \scriptsize,
  xleftmargin=.08\textwidth,
}

\lstdefinestyle{cfgstyle}{%
    comment = [l]{\#},
    basicstyle= \scriptsize,
    keywordstyle=\bfseries\em,
    keywords={,xori, andi, b, lw, sw, add, addi, ret_ra, addiu, jr, jal, copy, phi, mul, beq, retra, slti, b, blez, }
}

\newsavebox{\measurebox}

\makeatletter
\newcolumntype{"}{@{\hskip\tabcolsep\vrule width 1pt\hskip\tabcolsep}}
\makeatother

\begin{acronym}
\acro{CSP}{Constraint Satisfaction Problem}
\acro{COP}{Constraint Optimization Problem}
\acro{LNS}{Large Neighborhood Search}
\acro{DLNS}{Decomposition-based Large Neighborhood Search}
\acro{RS}{Random Search}
\acro{ATGP}{Automatic Generation of Architectural Tests}
\acro{ROP}{Return-Oriented Programming}
\acro{ISA}{Instruction Set Architecture}
\acro{IoT}{Internet of Things}
\acro{JOP}{Jump-Oriented Programming}
\acro{JIT-ROP}{Just-in-time Re\-turn-Oriented Programming}
\acro{HD}{Hamming Distance}
\acro{RD}{Register Hamming Distance}
\acro{LD}{Levenshtein Distance}
\acro{GD}{Gadget Distance}
\acro{CFI}{Control-Flow Integrity}
\acro{CFG}{Control-Flow Graph}
\acro{XoM}{Execute-Only Memory}
\acro{DivCon}{Diversity by Construction}
\acro{IR}{Intermediate Representation}
\acro{MIR}{Machine Intermediate Representation}
\acro{CP}{Constraint Programming}
\acro{FS}{Function Shuffling}
\acro{NFS}{No Function Shuffling}
\acro{CCITT}{International Telegraph and Telephone Consultative Committee}
\acro{DEP}{Data Execution Prevention}

\end{acronym}

\pgfplotstableset{
  begin table=\begin{longtable},
  end table=\end{longtable},
}

\jairheading{1}{2021}{1-15}{3/21}{}
\ShortHeadings{Constraint-based Diversification of JOP  Gadgets}
{Tsoupidi, Casta\~neda Lozano, \& Baudry}
\firstpageno{1}

\begin{document}
\title{Constraint-based Diversification of JOP  Gadgets}

\author{\name Authors}
\author{\name Rodothea Myrsini~Tsoupidi \email tsoupidi@kth.se \\
  \addr Royal Institute of Technology, KTH, \\ 
  Stockholm, Sweden 
  \AND
  \name Roberto~Casta\~neda Lozano \email roberto.castaneda@ed.ac.uk \\
  \addr University of Edinburgh, \\
  Edinburgh, United Kingdom
  \AND
  \name Benoit~Baudry \email baudry@kth.se \\
  \addr Royal Institute of Technology, KTH, \\ 
  Stockholm, Sweden 
}

\maketitle

\begin{abstract}

  Modern software deployment process produces software that is
  uniform, and hence vulnerable to large-scale code-reuse attacks,
  such as \emph{Jump-Oriented Programming (JOP)} attacks.
  \emph{Compiler-based diversification} improves the resilience
  and security of software systems by automatically generating
  different assembly code versions of a given program.
  Existing techniques are efficient but do not have a precise control
  over the quality, such as the code size or speed, of the generated
  code variants.

  This paper introduces \emph{Diversity by Construction (DivCon)}, a
  constraint-based compiler approach to soft\-ware diversification.
  Unlike previous approaches, DivCon allows users to control and adjust the
  conflicting goals of diversity and code quality. 
  A key enabler is the use of Large Neighborhood Search (LNS) to generate highly
  diverse assembly code efficiently.
  For larger problems, we propose a  combination of
  LNS with a structural decomposition of the problem.
  To further improve the diversification efficiency of DivCon against JOP
  attacks,
  we propose an application-specific distance measure tailored to the
  characteristics of JOP attacks.
  
  We evaluate DivCon with 20 functions from a popular benchmark suite
  for embedded systems.
  These experiments show that DivCon's combination of LNS and our
  application-specific distance measure generates binary programs that
  are highly resilient against JOP attacks (share between 0.15\% to 8\% of JOP gadgets) with an
  optimality gap of $\le$ 10\%.
  %
  %
  Our results confirm that there is a trade-off between the quality
  of each assembly code version and the diversity of the entire pool
  of versions.
  In particular, the experiments show that DivCon is able to generate
  binary programs that share a very small number of gadgets, while
  delivering near-optimal code.

  For constraint programming researchers and practitioners, this paper
  demonstrates that LNS is a valuable technique for finding diverse
  solutions.
  For security researchers and software engineers, DivCon extends the
  scope of compiler-based diversification to performance-critical and
  resource-constrained applications.

\end{abstract}

\renewcommand{\thelstlisting}{\arabic{lstlisting}}

\section{Introduction}
\label{sec:introduction}

Common software development practices, such as code reuse~\cite{Krueger1992}
and automatic updates, contribute
to the emergence of software monocultures~\cite{birman2009monoculture}.
While such monocultures facilitate software distribution, bug reporting,
and software authentication,
they also introduce serious risks related to the wide spreading of attacks
against all users that run identical software.

Embedded devices, such as controllers in cars or medical implants,
which manage sensitive and safety-critical data, are particularly
exposed to this class of
attacks~\shortcite{kornau2010return,bletsch_jump-oriented_2011}. Yet,
this type of software usually cannot afford expensive defense
mechanisms~\shortcite{salehi2019microguard}.

Software diversification is a method to mitigate the problems caused
by software monocultures, initially explored in the seminal work of
Cohen \cite{cohen1993operating} and Forrest
\cite{forrest1997building}.
Similarly to biodiversity, software diversification improves the
resilience and security of a software
system~\cite{baudryMultipleFacetsSoftware2015} by introducing diverse
variants of code in it.
Software diversification can be applied in different phases of the
software development cycle, i.e.\ during implementation, compilation,
loading, or execution~\shortcite{larsen_sok_2014}.
This paper is concerned with \emph{compiler-based} diversification,
which automatically generates different binary code versions from a
single source program.

Modern compilers do not merely aim to generate correct code, but also
code that is of high quality. There exists a variety of compilation
techniques to optimize code for speed or size
\shortcite{ashouri2018survey}.
However, there exist few compiler optimization techniques that target code diversification. 
These techniques are effective
at synthesizing diverse variants of  assembly code for one source program~\cite{larsen_sok_2014}. However, they do not have a precise control over other binary code quality metrics, such as speed or size.
These techniques (discussed in Section~\ref{sec:rel}) are either based on
randomizing heuristics or in high-level superoptimization methods that do not
capture accurately the quality of the generated code.

This paper introduces \ac{DivCon}, a compiler-based diversification approach
that allows users to control and adjust the conflicting goals of quality of each
code version and diversity among all versions.
\ac{DivCon} uses a \ac{CP}-based compiler backend to generate diverse solutions
corresponding to functionally equivalent program variants
according to an accurate code quality model.
The backend models the input program, the hardware architecture, and
the compiler transformations as a constraint problem, whose solutions
correspond to assembly code for the input program. The synthesis of
code diversity is motivated by \acf{JOP} attacks
\shortcite{checkoway_return-oriented_2010,bletsch_jump-oriented_2011} that
exploit the presence of certain binary code snippets, called JOP
gadgets, to craft an exploit. Our goal is to generate binary variants
that are functionally equivalent, yet do not have the same gadgets and
hence cannot be targeted by the exact same JOP attack.

The use of \ac{CP} makes it possible to 1) control the quality of the generated
solutions by constraining the objective function, 2) introduce
constraints tailored towards JOP gadgets, and 3)
apply search procedures that are particularly suitable for
diversification.
In particular, we propose to introduce  \ac{LNS}~\shortcite{shaw_using_1998}, a popular
metaheuristic in multiple application domains, to generate highly diverse
binaries.
For larger problems, we investigate a  combination of
\ac{LNS} with a structural decomposition of the problem.
Focusing on our application, \ac{DivCon} provides different distance measures
that trade diversity for scalability.

Our experiments compiling 14 functions from a popular embedded systems
suite to the MIPS32 architecture confirm that there is a trade-off
between code quality and diversity.
We demonstrate that \ac{DivCon} allows users to navigate
this space of near-optimal, diverse assembly code for a range of quality bounds.
We show that the Paretto front of optimal solutions synthesized by
\ac{DivCon} with LNS and a distance measure tailored against \ac{JOP}
attacks, naturally includes code variants with few common gadgets.
We  show that \ac{DivCon} is able to synthesize significantly
diverse variants while guaranteeing a code quality of 10\% within
optimality.
We further evaluate an additional set of six functions, which
  belong to the set of the 30\% largest functions of the benchmark suite, to investigate
  the scalability of \ac{DivCon}. 
%

For constraint programming researchers and practitioners, this paper
demonstrates that LNS is a valuable technique for finding diverse solutions.
For security researchers and software engineers, DivCon extends the scope of
compiler-based diversification to performance-critical and resource-constrained
applications, and provides a solid step towards secure-by-construction software.

To summarize, the main contributions of this paper are:
\begin{itemize}
\item the first \ac{CP}-based technique for compiler-based,
  quality-aware software diversification;
\item an experimental demonstration of the effectiveness of \ac{LNS}
  at generating highly diverse solutions efficiently;
\item the evaluation of DivCon on a wide set of benchmarks of different sizes, including large functions of up to 500 instructions;
\item a quantitative assessment of the technique to mitigate
  code-reuse attacks effectively while preserving high code quality; and
\item a publicly available tool for constraint-based software
  diversification\footnote{\url{ https://github.com/romits800/divcon}}.
\end{itemize}

This paper extends our previous work~\cite{tsoupidi2020constraint}.
We extend our investigation of LNS for code diversification with
\ac{DLNS} (Sections~\ref{ssec:algo}, \ref{ssec:evallns},
and~\ref{ssec:seval_algo}), a specific \ac{LNS}-based approach for
generating diverse solutions for larger programs.
We propose a new distance measure to explore the space of program
variants, which specifically targets JOP gadgets: \ac{GD}
(Sections~\ref{ssec:distances}, \ref{ssec:scale_dist},
and~\ref{ssec:gadget_dist}).
We perform a new set of experiments to compare the diversification
algorithms and the distance measures, with 19 new benchmark functions
up to 16 times larger than our previous dataset, providing new
insights on the scalability of our approach (Section~\ref{ssec:evallns}).
Finally, we add a case study on a voice compression application, which provides a
more complete picture on whole-program, multi-function diversification using
\ac{DivCon} (Section~\ref{ssec:casestudy}).

\section{Background}
\label{sec:background}

This section describes code-reuse attacks (Section~\ref{ssec:attacks}),
diversification approaches in \ac{CP} (Section~\ref{ssec:bakdivcp}), and
combinatorial compiler backends (Section~\ref{ssec:unison}).

\subsection{\ac{JOP} Attacks}
\label{ssec:attacks}

Code-reuse attacks take advantage of memory vulnerabilities, such as
buffer overflows, to reuse program legitimate code and repurpose it
for malicious usages.
More specifically, code-reuse attacks insert data into the program
memory to affect the control flow of the program.
Consequently, the original, valid code is executed but the modified
control flow triggers and executes code that is valid but unintended.

\ac{ROP}~\shortcite{shacham_geometry_2007}
is a code-reuse attack 
that combines different snippets from the original binary code
to form a Turing complete language for attackers. The building blocks of a
\ac{ROP} attack are the \emph{gadgets}: meta-instructions that consist of
one or multiple code snippets with specific semantics.
The original publication considers the x86 architecture and
the gadgets terminate with a \texttt{ret} instruction.
Later publications generalize \ac{ROP}, for different architectures
and in the absence of \texttt{ret} instructions, such as
\ac{JOP}~\shortcite{checkoway_return-oriented_2010,bletsch_jump-oriented_2011}.
This paper focuses on \ac{JOP} due to the
characteristics of MIPS32, but could be generalized to other
code-reuse attacks.
The code snippets for a \ac{JOP} attack
terminate with a branch instruction.  
Figure~\ref{lst:mips} shows a \ac{JOP} gadget found by the
\emph{ROPgadget} tool~\cite{ROPGadget2020} in a MIPS32 bi\-na\-ry.
Assuming that the attacker controls the stack, lines 2 and 3 load
attacker data in registers \$s2 and \$s4, respectively.  Then, line 4
jumps to the address of register \$t9.  The last instruction (line 5)
is placed in a delay slot and hence it is executed before the
jump~\cite{Sweetman2006}.  The semantics of this gadget depends on the
attack payload and might be to load a value to register \$s2 or \$s4.
Then, the program jumps to the next gadget that resides at the stack
address of \$t9.

\begin{figure}[t]
  \subfloat[Original gadget.]{%
    \label{lst:mips}
    \begin{minipage}[b]{.48\textwidth}
\begin{lstlisting}[style=mipsstyle]
0x9d001408: ...
0x9d00140c: lw    $s2, 4($sp)
0x9d001410: lw    $s4, 0($sp)
0x9d001414: jr    $t9
0x9d001418: addiu $sp, $sp, 16

\end{lstlisting}
      \vspace{-0.2cm}
    \end{minipage}
}
\subfloat[Diversified gadget.]{%
    \label{lst:mips2}
    \begin{minipage}[b]{.48\textwidth}
\begin{lstlisting}[style=mipsstyle]
0x9d001408: lw    $s2, 4($sp)
0x9d00140c: nop
0x9d001410: lw    $s4, 0($sp)
0x9d001414: jr    $t8
0x9d001418: addiu $sp, $sp, 16
\end{lstlisting}
      \vspace{-0.2cm}
    \end{minipage}
}
\caption{\label{fig:mips_example} Example gadget diversification in MIPS32 assembly code}
\end{figure}
  
Statically designed \ac{JOP} attacks use the absolute binary addresses
for installing the attack payload.
Hence, a simple change in the instruction schedule of the program as
in Figure~\ref{lst:mips2} prevents a \ac{JOP} attack designed for
Figure~\ref{lst:mips}.
An attacker that designs an attack based on the binary of the original
program assumes the presence of a gadget (Figure~\ref{lst:mips}) at
position 0x9d00140c.
However, in the diversified version, address 0x9d00140c does not start
with the initial \texttt{lw} instruction of Figure~\ref{lst:mips}, and
by the end of the execution of the gadget, register \$s2 does not
contain the attacker data.
Also, by assigning a different jump target register, \$t8, the next
target will not be the one expected by the attacker.
In this way, diversification can break the semantics of the gadget
and mitigate an attack against the diversified code.

\subsection{Attack Model}
We assume an attack model, where the attacker 1) knows the original C
code of the application, but 2) does not know the exact variant that
each user runs because we assume that each user runs a different
diversified version of the program, as suggested
by~\citeauthor{larsen_sok_2014}~\citeyear{larsen_sok_2014}.
%
Also, 3) we assume the existence of a memory corruption vulnerability
that enables a buffer overflow.
The defenses of the users include, \ac{DEP} (or $W\oplus X$), which
ensures that no writable memory ($W$) is executable ($X$) and vice
versa.
%
This ensures that the attacker is not able to execute code that is
directly inserted e.g.\ into the stack.

For more advanced attacks, like JIT-ROP
attacks~\cite{snow_just--time_2013}, we discuss later
(Section~\ref{ssec:advrop}) possible configurations using our
approach.

\subsection{Diversity in \acl{CP}}
\label{ssec:bakdivcp}
While typical \ac{CP} applications aim to discover either some solution or the
optimal solution, some applications require finding \textit{diverse} solutions
for various purposes.

\shortciteauthor{hebrard_finding_2005}~\citeyear{hebrard_finding_2005}
introduce the \textsc{MaxDiverse$k$Set} problem, which consists in
finding the most diverse set of $k$ solutions, and propose an exact
and an incremental algorithm for solving it.
The exact algorithm does not scale to a large number of solutions
\shortcite{van_hentenryck_constraint-based_2009,ingmar_modelling_2020}.
The incremental algorithm selects solutions iteratively by solving a
distance maximization problem.

\ac{ATGP} is an application of \ac{CP} that
 requires
generating many diverse solutions.
\shortciteauthor{van_hentenryck_constraint-based_2009}~\citeyear{van_hentenryck_constraint-based_2009}
model  \ac{ATGP}
as a \textsc{MaxDiverse$k$Set} problem and solve it
using the incremental algorithm of \shortciteauthor{hebrard_finding_2005}~\citeyear{hebrard_finding_2005}.
Due to the large number of
diverse solutions required (50-100), 
\shortciteauthor{van_hentenryck_constraint-based_2009}~\citeyear{van_hentenryck_constraint-based_2009}
replace the maximization step
with local search.

In software diversity, solution quality is of paramount importance.
In general, earlier \ac{CP} approaches to diversity are concerned with
satisfiability only.  An exception is the approach of
\citeauthor{petit_finding_2015}~\citeyear{petit_finding_2015}.
This approach modifies the objective function for assessing both solution
quality and solution diversity, but does not scale to the large number of
solutions required by software diversity.
\shortciteauthor{ingmar_modelling_2020}~\citeyear{ingmar_modelling_2020} propose a
generic framework for modeling diversity in \ac{CP}.
For tackling the
quality-diversity trade-off, they propose constraining the objective function
with the optimal (or best known) cost \textit{o}.
\ac{DivCon} applies this approach by allowing solutions $p\%$ worse
than \textit{o}, where $p$ is configurable.

\subsection{Compiler Optimization as a Combinatorial Problem } 
\label{ssec:unison}

A \acf{CSP} is a problem specification $P =\langle V,U, C\rangle$,
where $V$ are the problem variables,
$U$ is the domain of the variables, and $C$ the constraints among the variables.
A \acf{COP}, $P =\langle V,U, C, O \rangle$, consists of a \ac{CSP}
and an objective function $O$. The goal of a \ac{COP} is to find a solution
that optimizes $O$.

Compilers are programs that generate low-level assembly code, typically
optimized for \textit{speed} or \textit{size}, from higher-level source code.
A compilation process can be modeled as a \ac{COP} by letting $V$ be the
decisions taken during the translation, $C$ be the constraints that the
program semantics and the hardware resources impose, and $O$ be the cost of the
generated code.

\begin{figure} 
  \centering
\input{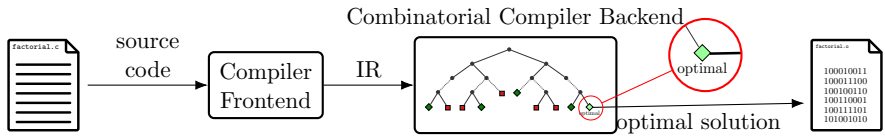}
\caption{\label{fig:unison}High-level view of a combinatorial compiler backend}
\end{figure}

Compiler backends generate low-level assembly code from an \ac{IR}, a program
representation that is independent of both the source and the target language.
Figure~\ref{fig:unison} shows the high-level view of a \emph{combinatorial}
compiler backend.
A combinatorial compiler backend takes as input the \ac{IR} of a program,
generates and solves a \ac{COP}, and outputs the optimized low-level assembly
code described by the solution to the \ac{COP}.

This paper assumes that programs at the \ac{IR} level are represented by
their \ac{CFG}.
A \ac{CFG} is a representation of the possible execution paths of a program,
where each node corresponds to a \emph{basic block} and edges correspond to
intra-block jumps.
A \emph{basic block}, in its turn, is a set of abstract instructions (hereafter
just \emph{instructions}) with no branches besides the end of the block.
Each instruction is associated with a set of operands characterizing its input
and output data.
Typical decision variables $V$ of a combinatorial compiler backend are the issue
cycle $c_i \in \mathbb{N}_0$ of each instruction $i$, the processor instruction
$m_i \in \mathbb{N}_0$ that implements each instruction $i$, and the processor
register $r_o \in \mathbb{N}_0$ assigned to each operand $o$.

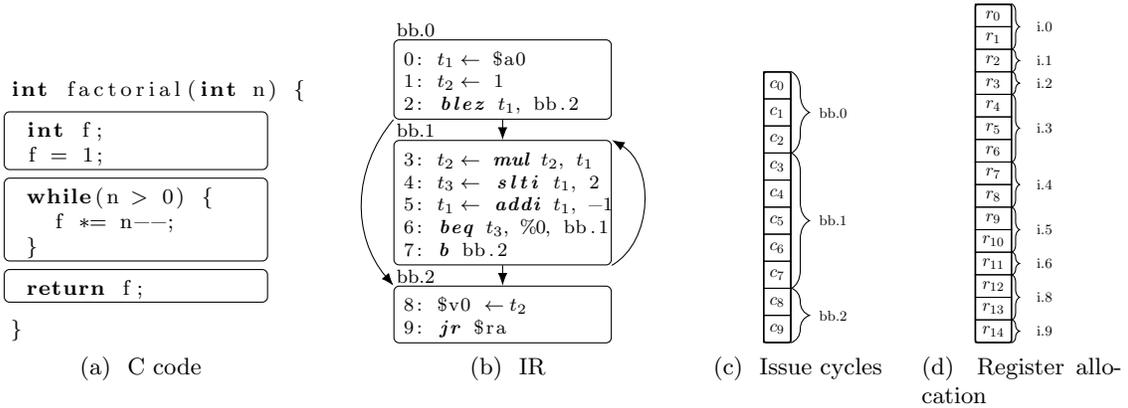
\begin{figure}[t]
  \newcommand{\dollar}{\mbox{\textdollar}}
  \centering
  \subfloat[\label{fig:fact} C code]{
    \begin{tikzpicture}[scale=1, transform shape,
    nst2/.style={
      inner xsep = -3pt,
      inner ysep = -6pt,
      align = left,
      minimum width = 4cm,
      text width = 4cm,
      anchor = south},
    nst/.style={
      draw,
      rounded corners=2pt,
      inner xsep = -3pt,
      inner ysep = -6pt,
      align = left,
      minimum width = 3.5cm,
      text width = 3.5cm,
      anchor = south}]

\node[nst2] (func1) {
  \begin{lstlisting}[style=cstyle]
int factorial(int n) {
\end{lstlisting}};

\node[nst, below=0.1cm of func1] (b0) {
  \begin{lstlisting}[mathescape=true, style=cstyle]
int f;
f = 1;
\end{lstlisting}};

\node[nst, below=0.1cm of b0] (b1) {
  \begin{lstlisting}[mathescape=true, style=cstyle]
while(n > 0) {
  f *= n--;
}
\end{lstlisting}};

\node[nst, below=0.1cm of b1] (b2) {
  \begin{lstlisting}[mathescape=true, style=cstyle]
return f;
\end{lstlisting}};

\node[nst2, below=0.1cm of b2] (func2) {
  \begin{lstlisting}[ style=cstyle]
}
\end{lstlisting}};

\end{tikzpicture}
  }
  \hfill
  \subfloat[\label{fig:factcfg} \ac{IR}]{
    \begin{tikzpicture}[scale=0.9, transform shape,
  nst/.style={
    draw,
    rounded corners=2pt,
    inner xsep = 3pt,
    inner ysep = -5pt,
    align = left,
    minimum width = 3cm,
    text width = 3cm,
    anchor = south}]

  \node[nst] (b0) {
      \begin{lstlisting}[mathescape=true, style=cfgstyle]
0: $t_1 \gets$ $\dollar$a0
1: $t_2 \gets$ 1
2: blez $t_1,$ bb.2
      \end{lstlisting}};

  \node[nst, below=0.3cm of b0] (b1) {
      \begin{lstlisting}[mathescape=true, style=cfgstyle]
3: $t_2 \gets$ mul $t_2,$ $t_1$
4: $t_3 \gets$ slti $t_1,$ 2
5: $t_1 \gets$ addi $t_1,$ -1
6: beq $t_3,$ %0$,$ bb.1
7: b bb.2
      \end{lstlisting}};

    \node[nst, below=0.3cm of b1] (b2) {
      \begin{lstlisting}[mathescape=true, style=cfgstyle]
8: $\dollar$v0 $\gets t_2$
9: jr $\dollar$ra
      \end{lstlisting}};

    \node[inner sep=0.4mm, anchor=south west] at (b0.north west) {\scriptsize{bb.0}};
    \node[inner sep=0.4mm, anchor=south west] at (b1.north west) {\scriptsize{bb.1}};
    \node[inner sep=0.4mm, anchor=south west] at (b2.north west) {\scriptsize{bb.2}};

    \draw[->, >={Latex}] (b0) -- (b1);
    \draw[->, >={Latex}] (b1) -- (b2);
    \draw[->, >={Latex}] (b0.south west) to [out=230, in=130]  (b2.north west);
    \draw[->, >={Latex}] (b1.south east) to [out=30, in=330]  (b1.north east);

\end{tikzpicture}
    
  }
  \hfill
  \subfloat[\label{fig:cyc} Issue cycles]{
    \begin{tikzpicture}[scale=0.6, transform shape,
  nst/.style={rectangle, minimum width = 0.6cm, minimum height = 0.6cm, draw, anchor=north west},
  ast/.style={decorate,decoration={brace,amplitude=7pt},xshift=0pt,yshift=-4pt},
  mst/.style={black, midway, right, xshift=14pt}]

  \node[rectangle,
    minimum width =  0.6cm,
    minimum height = 6cm,
    draw,
    anchor=north west,
    rounded corners=1pt,
    semithick] at (0,0) {};

  \foreach \x in {0,...,9} {
    \node[nst] (n\x) at (0,-0.6*\x) {$c_\x$}; 
  }

\draw [ast] (n0.north east) -- (n2.south east) node [mst] {\footnotesize bb.0};
\draw [ast] (n3.north east) -- (n7.south east) node [mst] {\footnotesize bb.1};
\draw [ast] (n8.north east) -- (n9.south east) node [mst] {\footnotesize bb.2};
\node [right = 40pt of n1] {};
\node [left = 20pt of n1] {};
  
\end{tikzpicture}
    
  }
  \hfill
  \subfloat[\label{fig:reg} Register allocation]{
    \begin{tikzpicture}[scale=0.6, transform shape,
  nst/.style={rectangle, minimum width = 0.8cm, minimum height = 0.5cm, draw, anchor=north west},
  ast/.style={decorate,decoration={brace,amplitude=3pt},xshift=0pt,yshift=-4pt},
  mst/.style={black, midway, right, xshift=12pt}]

  \node[rectangle,
    minimum width =  0.8cm,
    minimum height = 7.5cm,
    draw,
    anchor=north west,
    rounded corners=1pt,
    semithick] at (0,0) {};

  \foreach \x in {0,...,14} {
    \node[nst] (n\x) at (0,-0.5*\x) {$r_{\x}$}; 
  }

  \draw [ast] (n0.north east) -- (n1.south east) node [mst] {\footnotesize i.0};
  \draw [ast] (n2.north east) -- (n2.south east) node [mst] {\footnotesize i.1};
  \draw [ast] (n3.north east) -- (n3.south east) node [mst] {\footnotesize i.2};
  \draw [ast] (n4.north east) -- (n6.south east) node [mst] {\footnotesize i.3};
  \draw [ast] (n7.north east) -- (n8.south east) node [mst] {\footnotesize i.4};
  \draw [ast] (n9.north east) -- (n10.south east) node [mst] {\footnotesize i.5};
  \draw [ast] (n11.north east) -- (n11.south east) node [mst] {\footnotesize i.6};
  \draw [ast] (n12.north east) -- (n13.south east) node [mst] {\footnotesize i.8};
  \draw [ast] (n14.north east) -- (n14.south east) node [mst] {\footnotesize i.9};
\node [right = 50pt of n1] {};
\node [left = 20pt of n1] {};
  
\end{tikzpicture}
    
  }
  \caption{\label{fig:factorial} Factorial function example}
\end{figure}

Figure~\ref{fig:fact} shows an implementation of the factorial
function in C where each basic block is highlighted.
Figure~\ref{fig:factcfg} shows the
\ac{IR} of the program.
The example \ac{IR} contains 10 instructions in three basic blocks: bb.0, bb.1,
and bb.2.
Basic block bb.0 corresponds to initializations, where \texttt{\$a0}
holds the function argument \texttt{n} and $t_1$ corresponds to
variable \texttt{f}.
bb.1 computes the factorial in a loop by accumulating the result in $t_2$.
bb.2 stores the result to \texttt{\$v0} and returns.
Some instructions in the example are interdependent, which leads to
serialization of the instruction schedule.
For example, \textbf{\textit{beq}} (6) consumes data ($t_3$) defined by
\textbf{\textit{slti}} (4) and hence needs to be scheduled later.
Instruction dependencies limit the amount of possible assembly code versions and
may restrict diversity significantly. 
Finally, Figure~\ref{fig:cyc} shows the arrangement of the issue-cycle
variables in the constraint model used by the combinatorial compiler
backend. 
Similarly, Figure~\ref{fig:reg} shows the arrangement of the register
variables.

The \ac{CFG} representation of a program offers a natural
decomposition of the \ac{COP} into subproblems, each consisting of a
basic block.  This partitioning requires first solving the
\textit{global} problem that assigns registers to the program
variables that are live (active) through different basic blocks
\shortcite{lozano_constraint-based_2012}.
For example, in Figure~\ref{fig:factcfg}, the global problem has to
assign a register to $t_1$ because both bb.0 and bb.1 use it.
Subsequently, it is possible to solve the \ac{COP} by optimizing
each of the \textit{local} problems (for every basic block),
independently.

\ac{DivCon} aims at mitigating code-reuse attacks.
Therefore, \ac{DivCon} considers the order of the instructions and the
assignment of registers to their operands in the final binary, which
directly affects the feasibility of code-reuse attacks (see
Figures~\ref{lst:mips} and \ref{lst:mips2}).
For this reason, the diversification model uses the issue-cycle
sequence of instructions, $c = \{c_0,c_1,..., c_n\}$, and the register
allocation, $r = \{r_0,r_1,...,r_n\}$, to characterize the diversity
among different solutions.

\section{DivCon}
\label{sec:approach}

\begin{figure}[t]
  \centering
  \input{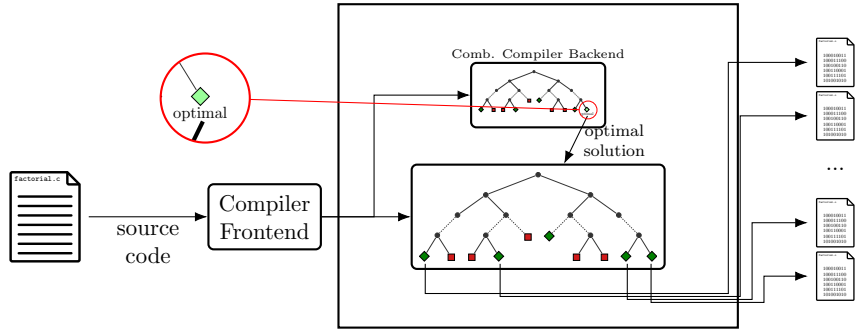}
  \caption{\label{fig:divcon} High-level view of \ac{DivCon}}
\end{figure}


This section introduces \ac{DivCon}, a software diversification method
that uses a combinatorial compiler backend to generate program variants.
Figure~\ref{fig:divcon} shows a high-level view 
of the diversification process.
\ac{DivCon} uses 1) the optimal solution 
(see Definition~\ref{def:opt_sol}) to start the \textit{search} for
diversification and 2) the cost of the optimal solution to
restrict the variants within a maximum gap from the optimal.
Subsequently, \ac{DivCon} generates a
number of solutions to the \ac{CSP} that correspond
to diverse program variants.

The rest of this section describes the diversification
approach of \ac{DivCon}.
Section~\ref{ssec:problem}
formulates the diversification problem
in terms of the constraint
model of a combinatorial compiler backend,
Section~\ref{ssec:algo} defines the proposed diversification algorithms,
Section~\ref{ssec:distances} defines  the 
distance measures, and finally, Section~\ref{ssec:search}
describes the search strategy for generating program variants.

\subsection{Problem Description}
\label{ssec:problem}


In this section, we will define the program diversification problem and stress
important concepts that we will use later in the evaluation part (Section~\ref{sec:evaluation}).
Let $P = \langle V,U,C\rangle$ be the compiler backend \ac{CSP} for
the program under compilation and $O$ the objective function of the
\ac{COP}, $\langle V,U,C,O\rangle$.

\begin{definition} 
\textbf{Optimal solution} is the solution $y_{opt}\in sol(P)$ that the
combinatorial compiler backend (see Section~\ref{ssec:unison}) returns
and for which $O(y_{opt}) = o$.
\label{def:opt_sol}
\end{definition}

We then define the \textit{optimality gap}
as follows: 

\begin{definition}
\textbf{Optimality gap} is the ratio, $p \in \mathbb{R}_{\geq 0}$, that 
constrains the optimization function, such that  $\forall s \in sol(P) \, . \,
O(s) \leq (1+p) \cdot o $.
\label{def:opt_gap}
\end{definition}

We define the \textit{distance} function (three such functions are
defined in Section~\ref{ssec:distances}) as follows:

\begin{definition}
\textbf{Distance} $\delta(s_1,s_2)$ is a function that measures the
distance between two solutions of P, $s_1,s_2 \in sol(P)$.
\label{def:deltas}
\end{definition}

Let parameter $h \in \mathbb{N}$ be
the minimum allowed pairwise distance between two generated solutions.
Our problem is to find a subset of the solutions to the \ac{CSP}, $S \subseteq sol(P)$,
such that:
\begin{equation}
\forall s_1, s_2 \in S\, . \, s_1 \neq s_2 
\implies \delta(s_1,s_2) \ge h\text{ and }\forall s \in S \, . \,
O(s) \leq (1+p) \cdot o
\end{equation}

To solve the above problem, this paper proposes two \ac{LNS}-based
incremental algorithms defined in Section~\ref{ssec:algo}.
\ac{LNS} is a metaheuristic that allows searching for solutions in
large parts of the search tree.
This property makes \ac{LNS} a good candidate for generating a
large number of diverse solutions. 
To guarantee that the new variants are sufficiently different from
each other, we define three distance measures
(Section~\ref{ssec:distances}) that quantify the concept of program
difference for our application.




\subsection{Diversification Algorithms}
\label{ssec:algo}
This section presents two \ac{LNS}-based algorithms for the generation
of a large number of solutions for software diversification.
The first algorithm (Algorithm~\ref{alg:search}), referred to as
simply \ac{LNS}, solves the problem monolithically using an
\ac{LNS}-based approach, whereas the second algorithm, \ac{DLNS}
(Algorithm~\ref{alg:decomp}), decomposes the problem into subproblems
and uses \ac{LNS} to diversify each of these subproblems independently
and in parallel.
The final solutions are then composed by randomly combining the
solutions of the subproblems. 

\paragraph{\ac{LNS} Algorithm.}
Algorithm~\ref{alg:search} presents a monolithic \ac{LNS}-based
diversification algorithm.
It starts with the optimal solution $y_{opt}$ (line 3).
Subsequently, the algorithm adds a distance constraint for $y_{opt}$
and the optimality constraint with $o = O(y_{opt})$ (line 4).
While the termination condition is not fulfilled (line 5), the
algorithm uses \ac{LNS} as described in Section~\ref{ssec:search} to
find the next solution $y$ (line 6), adds the next solution to the
solution set $S$ (line 7), and updates the distance constraints based
on the latest solution (line 8).
When the termination condition is satisfied, the algorithm returns the
set of solutions $S$ corresponding to diversified assembly code
variants (line 9).

In our experience, our application does not require large values of
$h$ because even small distance between variants breaks gadgets (see
Figure~\ref{fig:mips_example}).
An alternative algorithm that may improve Algorithm~\ref{alg:search}
for larger values of $h$, is replacing \texttt{solve$_{LNS}$} on line
6 and the constraint update on line 8 with an \ac{LNS} maximization
step that returns a solution by iteratively improving its pairwise
distance with all current solutions in $S$ until reaching the value of
$h$.

\begin{algorithm}[caption={Incremental algorithm for generating diverse solutions},
    label={alg:search}]
  function solve_lns($y_{opt}$, $\langle V,U, C\rangle$)
  begin
    S $\gets$ {$y_{opt}$}, y $\gets y_{opt}$,
    $C' \gets C \,\cup \{\delta(y_{opt}) \ge h$, $O(V) \leq (1 + p)\cdot o\}$
    while not term_cond() // e.g. |S| > k $\lor$ time_limit()
      y $\gets$ solve$_{LNS}$(relax(y), $\langle V,U, C'\rangle$)
      S $\gets$ S $\cup$ {y}
      $C' \gets C'\,\cup \{\delta(y,s) \ge h$ | $\forall s \in sol(\langle V,U,C'\rangle)$ $\}$
    return S
  end
\end{algorithm}

\paragraph{Decomposition Algorithm.}
This section presents \ac{DLNS} (Algorithm~\ref{alg:decomp}),
an \ac{LNS}-based algorithm that uses decomposition
to enable diversification of large functions.
To enable this, the algorithm divides the problem into a
\textit{global} problem and a set of \textit{local} subproblems, one for each basic
block of the function.

Algorithm~\ref{alg:decomp}
starts by adding the optimal solution to the set of solutions (line 3)
and continues by adding the optimality constraints (line 4).  While
the termination condition is not satisfied, the algorithm solves
the \textit{global} problem  (line 7). 
After finding a global solution, the
algorithm solves the \textit{local} problems, i.e. for each basic
block $b\in B$, in parallel and generates a number of \textit{local}
solutions for each basic block (lines 9 and 10).
Then, the algorithm combines one randomly selected solution for each
basic block (line 13).
This combined solution may be invalid (line 14), due to, for example,
exceeded \textit{cost}.
In case the solution is valid (line 14),
%
%
the algorithm adds this solution to the set of
solutions $S$ (line 15) and, finally, adds a diversity constraint to
the problem (line 16). 

\begin{algorithm}[caption={Decomposition-based incremental algorithm for generating diverse solutions},
    label={alg:decomp}]
  function solve_decomp_lns($y_{opt}$, $\langle V,U, C\rangle$)
  begin
    S $\gets$ {$y_{opt}$}, y $\gets y_{opt}$,
    $C' \gets C \,\cup \{\delta(y_{opt}) \ge h$, $O(V) \leq (1 + p)\cdot o\}$
    while not term_cond() // e.g. |S| > k $\lor$ time_limit()
      // Find partial solution
      y $\gets$ psolve$_{LNS}$(relax(y), $\langle V,U, C'\rangle$)
      // Solve local problems
      for b $\in$ B
          S$_b$ $\gets$ spawn solve_lns(y$_{b}$, $\langle V_b,U_b, C'_b\rangle$)
      // Select solutions
      for |S$_1$ $\times$ S$_2$ $\times$ ... $\times$ S$_b$|
          y $\gets$ combine($\forall$ b $\in$ B.$\exists$ y$_b$ $\in$ S$_b$. y$_b$, $\langle V,U, C'\rangle$)
          if valid(y):
              S $\gets$ S $\cup$ {y}
              $C' \gets C'\,\cup \{$ $\delta($y$, s) \ge h $ | $\forall s \in sol(\langle V,U,C'\rangle)$ $\}$
  end
\end{algorithm}

\paragraph{Example.}
Figure~\ref{fig:mips_variants} shows two MIPS32 variants of the
factorial example (Figure~\ref{fig:factorial}), which correspond to two
solutions of \ac{DivCon}.
The variants differ in two aspects: first, the \textbf{\textit{beqz}}
instruction is issued one cycle later in Figure~\ref{lst:fact2} than
in Figure~\ref{lst:fact1}, and second, the temporary variable $t_3$
(see Figure~\ref{fig:factorial}) is assigned to different MIPS32
registers (\texttt{\$t0} and \texttt{\$t1}).
\ac{LNS} diversifies the function that consists of three basic
blocks by finding different solutions that assign values
to the registers and the instruction schedule simultaneously.
\ac{DLNS} solves first the global problem by assigning registers to the
temporary variables that are live across multiple basic blocks ($t_1$ and
$t_2$) and then assigns the issue schedule and the rest of the registers for
each basic block, independently and possibly in parallel.
The diversified variants in Figure~\ref{fig:mips_variants} serve
presentation purposes.
Figure~\ref{fig:mips_example_ext} in Appendix~\ref{sec:appb}
presents a more elaborated example of two diversified functions.%

\begin{figure}[b]
  \subfloat[Variant 1]{%
    \label{lst:fact1}
  \begin{minipage}[t]{0.38\textwidth} %
\begin{lstlisting}[style=mipsstyle, firstline=20,lastline=27]
bb.0: blez  $a0, bb.2
      addiu $v0, $zero, 1
bb.1: mul   $v0, $v0, $a0
      slti  $t0, $a0, 2
      beqz  $t0, bb.1
      addi  $a0, $a0, -1
bb.2: jr    $ra
      nop
\end{lstlisting}
    \vspace{-0.2cm}
    \
  \end{minipage}}
  \hfill
  \subfloat[Variant 2]{%
    \label{lst:fact2}
  \begin{minipage}[t]{0.38\textwidth}
\begin{lstlisting}[style=mipsstyle, firstline=20,lastline=28]
bb.0: blez  $a0, bb.2
      addiu $v0, $zero, 1
bb.1: mul   $v0, $v0, $a0
      slti  $t1, $a0, 2
      nop
      beqz  $t1, bb.1
      addi  $a0, $a0, -1
bb.2: jr    $ra
      nop
\end{lstlisting}
    \vspace{-0.2cm}
  \end{minipage}}
  \caption{\label{fig:mips_variants} Two MIPS32 variants of the
factorial example in Figure~\ref{fig:factorial}}
\end{figure}

\subsection{Distance Measures}
\label{ssec:distances}
This section
defines three alternative 
distance measures:
\acf{HD}, \acf{LD},
and \acf{GD}.
\ac{HD} and \ac{LD} operate on the schedule of the
instructions, i.e.\ the order in which the instructions are issued
in the CPU, whereas \ac{GD} operates on both the instruction
schedule and the register allocation, i.e.\ the hardware register
for each operand.
Early experimental results  that we have performed
have shown that
diversifying register allocation is less effective than diversifying
the instruction schedule against code-reuse attacks.
%
However,
restricting register allocation diversity to the instructions very
near a branch instruction (a key component of a \ac{JOP} gadget),
improves \ac{DivCon}'s gadget diversification effectiveness.


\paragraph{\acf{HD}.} \ac{HD}
is the Hamming distance~\cite{hamming1950error} between the
issue-cycle variables of two  solutions.
Given two solutions $s, s' \in sol(P)$:
\begin{equation}
  \delta_{HD} (s,s') = \displaystyle\sum_{i=0}^{n} (s(c_i) \neq s'(c_i)),
  \label{eq:hdist}
\end{equation}
\noindent
where $n$ is the maximum number of instructions.

Consider Figure~\ref{lst:mips2}, a diversified version of the gadget
in Figure~\ref{lst:mips}.
The only instruction that differs from
Figure~\ref{lst:mips} is the instruction at line 1 that is
issued one cycle before. The two examples
have a \ac{HD} of one,
which in this case is enough for breaking
the functionality of the original gadget (see Section~\ref{ssec:attacks}).

\paragraph{\acf{LD}.}
\acl{LD} (or edit distance) measures the minimum number of edits, i.e.\
insertions, deletions, and replacements, that are necessary
for transforming one instruction schedule to another.
Compared to \ac{HD}, 
which considers only \textit{replacements},
\ac{LD} also considers \textit{insertions} and \textit{deletions}.
To understand this effect, consider Figure~\ref{fig:mips_variants}.
The two gadgets differ only by one \texttt{nop}
operation but \ac{HD} gives a distance
of three, whereas \ac{LD} gives one,
which is more accurate.
\ac{LD} takes ordered vectors as input, and thus requires an ordered
representation (as opposed to a detailed schedule) of the instructions.
Therefore, \ac{LD} uses vector $c^{-1}= channel(c)$, a sequence of instructions
ordered by their issue cycle.
Given two solutions $s, s' \in sol(P)$:
\begin{equation}
  \delta_{LD} (s,s') = \texttt{levenshtein\_distance}(s(c^{-1}),s'(c^{-1})), 
  \label{eq:levdist}
\end{equation}
\noindent
where \texttt{levenshtein\_distance} is the Wagner–Fischer algorithm~\cite{wagner1974string}
with time complexity 
$O(nm)$, where $n$ and $m$ are the lengths of the two sequences.

\paragraph{\acf{GD}.}
\ac{GD} is an  application-specific distance measure targeting
\ac{JOP} gadgets that we propose in this paper.
\ac{GD} operates on both register allocation and instruction scheduling
focusing on the instructions preceding  a branch instruction because \ac{JOP} gadgets
terminate with a branch instruction.
Here, the set of branch instructions, $B$, consists of all indirect
\textit{jump} or \textit{call} instructions (e.g.\ line 7 in Figure~\ref{lst:fact1}).
A gadget may also use a direct jump (e.g.\ line 5 in
Figure~\ref{lst:fact1}), however, the majority of gadgets require
control over the jump target, which is not possible with direct jumps.
\ac{GD} uses two configuration parameters, $n_c$ and $n_r$.
Parameter $n_c$ denotes the number of instructions before each branch,
$br\in B$, that the issue cycle
of two variants may differ.
Similarly, parameter $n_r$ denotes the
number of instructions preceding a branch
of two variants that the register assignment of
the instruction operands may differ.
Consider Figure~\ref{lst:mips2}.
The two gadgets differ by one \texttt{nop} instruction
and a different register at instruction 4.
Then, the \ac{GD} distance is two, given $n_c=3$ and $n_r=0$. 
%

Given two solutions $s, s' \in sol(P)$, the partial distance
$\delta_{PGD}^{n_r,n_c}$ on branch $br\in B$ is :
\begin{equation}
  \delta_{PGD}^{n_r,n_c}(s,s',br) = \displaystyle\sum_{i=0}^{N_i} \left(f(s, n_c, i, br)(s(c_i) \neq s'(c_i)) +
  \displaystyle\sum_{p\in ps(c_i)} f(s, n_r, i, br)(s(r_p) \neq s'(r_p)) \right),
  \label{eq:hdist}
\end{equation}

\noindent where $N_i$ is the number of instructions, $ps(c_i)$ is the
set of operands in instruction $i$, and $f(s,n,i,br)$ is a function
that takes four inputs, i) one solution $s\in S$, ii) a natural number that
corresponds to the allowed distance of an instruction $i$ from a
branch instruction $br$, iii) instruction $i$,
and iv) branch instruction $br$. The definition of $f$ is as follows:

\begin{equation}
  f(s,n, i,br) =  \begin{cases}
    1, & s(c_{br}) - s(c_i) \in [0, n]\\
    0, & \text{ otherwise}\\
    \end{cases}.
  \label{eq:hdist}
\end{equation}

Finally, given two solutions $s, s' \in sol(P)$, the \acl{GD}
$\delta_{GD}^{n_r,n_c}$ is defined as:

\begin{equation}
  \delta_{GD}^{n_r,n_c}(s,s') = \displaystyle\sum_{br\in B} \left(
  \delta_{PGD}^{n_r,n_c}(s,s',br) \right).
  \label{eq:hdist}
\end{equation}

Note that in Algorithm~\ref{alg:search} and Algorithm~\ref{alg:decomp},
\ac{GD} will result in a
number of constraints equal to the number of branches in $B$ plus one.

\subsection{Search}
\label{ssec:search}

Unlike previous CP approaches to diversity, \ac{DivCon} employs
\acf{LNS}~\cite{shaw_using_1998} for diversification.  \ac{LNS} is a
metaheuristic that defines a neighborhood, in which \textit{search}
looks for better solutions, or, in our case, different solutions.  The
definition of the neighborhood is through a \textit{destroy} and a
\textit{repair} function.  The \textit{destroy} function unassigns a
subset of the variables in a given solution and the \textit{repair}
function finds a new solution by assigning new values to the
\emph{destroyed} variables.

In \ac{DivCon}, the algorithm starts with the optimal solution
(Definition~\ref{def:opt_sol}) of the combinatorial compiler backend.
Subsequently, it destroys a part of the variables and continues with
the model's branching strategy to find the next solution, applying a
restart after a given number of failures.
\ac{LNS} uses the concept of \textit{neighborhood}, i.e.\ the variables that
\ac{LNS} may destroy at every restart.
To improve  diversity, the neighborhood for \ac{DivCon} consists of
all decision variables, i.e.\  the issue cycles $c$, the instruction implementations $m$,
and the registers $r$.
Furthermore, \ac{LNS} depends
on a \textit{branching strategy} to
guide the \textit{repair} search.
To improve security and allow \ac{LNS} to select diverse paths after every
restart, \ac{DivCon} employs a random variable-value selection branching
strategy as described in Table~\ref{tab:random}.

\begin{table}
  \caption{\label{tab:randomcloriginal} \textsc{Original} and \textsc{Random} branching strategies}
  \centering
  \subfloat[\label{tab:cloriginal} \textsc{Original} branching strategy]{
    {\footnotesize
      \begin{tabular}{|c|>{\centering}m{2cm}|c|}
        \hline
        Variable & Var. Selection & Value Selection \\\hline 
        $c_i$ &  in order & min.~val first \\
        $m_i$ &  in order & min.~val first \\
        $r_o$ &  in order & randomly \\\hline
      \end{tabular}
    }
  }\hfill
  \subfloat[\label{tab:random} \textsc{Random} branching strategy]{
    {\footnotesize
      \begin{tabular}{|c|>{\centering}m{2cm}|c|}
        \hline
        Variable & Var. Selection & Value Selection \\\hline 
        $c_i$ & randomly  &  randomly  \\
        $m_i$ & randomly  &  randomly \\
        $r_o$ & randomly  &  randomly  \\\hline
      \end{tabular}
    }
  }
\end{table}


\section{Evaluation}
\label{sec:evaluation}


This section evaluates \ac{DivCon} experimentally.
For simplicity, the section uses the acronyms \ac{LNS} and \ac{DLNS}
to refer to the specific application
of Algorithms~\ref{alg:search} and~\ref{alg:decomp} in \ac{DivCon}.
The diversification effectiveness
and the scalability of \ac{DivCon} depend on three main dimensions:
\begin{itemize}
\item \textbf{Optimality gap} (see Definition~\ref{def:opt_gap}),
  which relaxes the optimization  function. 
  Here, we evaluate the diversification effectiveness and scalability for four different values of $p$,
  $0\%$, $5\%$, $10\%$, and $20\%$
\item \textbf{Diversification algorithm}. We compare our two proposed
  algorithms, \ac{LNS} (Algorithm~\ref{alg:search}) and \ac{DLNS}
  (Algorithm~\ref{alg:decomp}) with \ac{RS} and
  incremental
  \textsc{MaxDiverse$k$Set}~\shortcite{hebrard_finding_2005}.
  \ac{RS} uses the branching strategy of
  Table~\ref{tab:random}.
  For \textsc{MaxDiverse$k$Set}, the first solution corresponds to the
  optimal solution (see Definition~\ref{def:opt_sol}) and the
  maximization step uses the branching strategy of
  Table~\ref{tab:cloriginal}.  
\item \textbf{Distance measure}. We compare four distance measures (Section~\ref{ssec:distances}), 
  \ac{HD}, $\delta_{HD}$, \ac{LD}, $\delta_{LD}$,
  and two configurations of \ac{GD} for different values of parameters
  $n_r$ and $n_c$ (see Section~\ref{ssec:distances}), $\delta_{LD}^{0,2}$
  and $\delta_{LD}^{0,8}$. The two parameters control the number of instructions
  preceding a branch that differ among different solutions.  
  The smaller these parameters are, the higher the chance of breaking
  a larger number of \ac{JOP} gadgets, 
  given that all gadgets end with a branch instruction.
\end{itemize}

The output of DivCon is a set of diverse binary variants. To evaluate the
diversification effectiveness of each approach, we compare the generated 
binaries using the following three measures:

\begin{itemize}
    \item \textbf{Code diversity}, which measures the pairwise distance of the final binaries using 
    the same distance that was used for diversification.
    The definition is in Equation~\ref{eq:pairwise}.
    \item \textbf{Gadget diversity}, which measures the rate of gadgets that \ac{DivCon} 
    diversifies successfully (see Section~\ref{ssec:seval_algo}).
  \item \textbf{Scalability}, which is related to the number of variants generated
    within a fixed time budget or the total time required
    to generate the maximum number of variants.
\end{itemize}

The six research questions below investigate the influence of the \textbf{optimality gap},
 \textbf{diversification algorithm}, \textbf{distance measure}, and \textbf{program scope}
with respect to our three diversity measures.
 
\begin{itemize}
\item RQ1. How effective are our two novel diversification algorithms?  Here, we compare \ac{LNS}
  and \ac{DLNS} with state-of-the-art diversification algorithm,
  with respect to their ability to generate binary code that
  is as diverse as possible.
  This question evaluates the \textbf{code diversity} of
  \ac{DivCon} for the different \textbf{diversification algorithms}.
\item RQ2. What is the scalability of the distance measures
  when generating multiple program variants?
  Here, we evaluate which of the distance measures
  is the most appropriate for software diversification.
  This question evaluates the \textbf{scalability} of
  \ac{DivCon} for the different \textbf{distance measures}.
\item RQ3. How effective is \ac{DivCon} using 
  \ac{LNS} and \ac{DLNS} at mitigating \ac{JOP} attacks? 
  In this part, we evaluate which method is the most
  effective against \ac{JOP} attacks by comparing the rate of shared
  gadgets among the generated solutions.
  This question evaluates the \textbf{gadget diversity} of
  \ac{DivCon} for the different \textbf{diversification algorithms}.
\item RQ4. How effective is \ac{DivCon} using different distance measures
  against \ac{JOP} attacks?
  Here, we evaluate the effectiveness of \ac{DivCon}
  using four different distance measures against \ac{JOP} attacks.
  This question evaluates the \textbf{gadget diversity} of
  \ac{DivCon} for the different \textbf{distance measures}.
\item RQ5. How does code quality affect the effectiveness of \ac{LNS}
  against \ac{JOP} attacks using an application-specific distance measure?
  Here, we evaluate the effect of code quality on the effectiveness
  of \ac{DivCon} at mitigating \ac{JOP} attacks.
  This question evaluates the \textbf{gadget diversity} of
  \ac{DivCon} for the different \textbf{optimality gaps}.
\item RQ6. What is the effect of function diversification
  with DivCon at the application level?
  Here, we evaluate the effect of diversification using  \ac{DivCon}
  with a voice compression case study.
  This question evaluates the \textbf{gadget diversity} of
  \ac{DivCon} in a compiled whole-program binary consisting of multiple functions.
\end{itemize}



\subsection{Experimental Setup}

The following paragraphs describe the experimental setup for the
evaluation of \ac{DivCon}.

\paragraph{Implementation.}
\ac{DivCon} is implemented as an extension of Unison~\cite{lozano_combinatorial_2019},
and is available online\footnote{\url{https://github.com/romits800/divcon}}.
Unison implements two backend transformations: instruction scheduling
and register allocation.
As part of register allocation, Unison captures many
interdependent transformations such as spilling, register
assignment, coalescing, load-store optimization, register packing,
live range splitting, rematerialization, and
multi-allocation~\cite{lozano_combinatorial_2019}.
Unison models two objective functions for code quality, \textit{speed}
and \textit{code size}.
This evaluation uses the \textit{speed} objective function, which
considers statically derived basic-block frequencies and the
execution time of each basic block that depends on the shared
resources, the instruction issue cycles, and the instruction
latencies.
These execution times and latencies were based on
a generic MIPS32 model of LLVM~\cite{lozano_combinatorial_2019}.
\ac{DivCon} relies on Unison's solver portfolio that includes Gecode
v6.2~\cite{Gecode2020} and Chuffed v0.10.3~\cite{Chu2011} to find optimal
binary programs.
We use Gecode v6.2 for automatic diversification because
Gecode provides an interface for customizing \textit{search}.
%
The LLVM compiler~\cite{Lattner2004} is used as a front-end and
\ac{IR}-level optimizer, as well as for the final emission of
assembly code.
\ac{DivCon} operates on the \ac{MIR}\footnote{Machine Intermediate Representation: \url{https://www.llvm.org/docs/MIRLangRef.html}} level of LLVM.

\paragraph{Benchmark functions and platform.}
We evaluate the ability of \ac{DivCon} to generate program variants
with 20 functions sampled randomly from
MediaBench\footnote{A later version of MediaBench,
    MediabBench II was not complete by the time we are writing this
    paper.}~\shortcite{Lee1997}.
This benchmark suite is widely employed in embedded systems research.
We select two sets of benchmarks.
The first set consists of 14 functions ranging from 10 to 100
\ac{MIR} instructions with a median of 58 instructions.
The second set consists of six functions ranging between
100 and 1000 lines of \ac{MIR} instructions.
Functions with size below 100 \ac{MIR} instructions compose the 65\%
of the functions in MediaBench, whereas functions with size less than
500 \ac{MIR} instruction compose the 93\%, and those with size less
than 1000 \ac{MIR} instructions compose the 97\% of the functions in
MediaBench.

Smaller functions in the first set allow the evaluation of all
algorithms and distance measures regardless of their computational
cost, whereas larger functions challenge our diversification
algorithms.
Table~\ref{tab:mediabench} lists the ID, application, function name,
the number of basic blocks, and the number of \ac{MIR} instructions of each
sampled function.
For evaluating the scalability of DivCon, we perform an
additional experiment consisting of the second set of
functions.
Table~\ref{tab:benchmarks_large} describes these additional
benchmarks.  
These benchmarks are used only for evaluating the scalability of \ac{DivCon}
due to time constraints.

Furthermore, for evaluating the effectiveness of our approach at the
application level, we perform a case study of one of application from
MediaBench, G.721.
This application consists of functions that we present in
Table~\ref{tab:benchmarks_g721}.

The functions are compiled to MIPS32 assembly code, a popular
architecture within embedded systems and the security-critical
\acl{IoT}~\shortcite{alaba_internet_2017}.



\paragraph{Host platform.}
All experiments run on
an Intel%
\textsuperscript{\textregistered}%
Core\texttrademark i9-9920X processor at 3.50GHz with 64GB of RAM
running Debian GNU/Linux 10 (buster).
Each experiment runs for 15 random seeds.
The aggregated results of the evaluation (RQ1)
show the mean value and the standard deviation
for the maximum number of generated variants, where
at least five seeds are able to terminate within a time limit. 
  For the smaller benchmarks (Table~\ref{tab:mediabench}),
  we have 10GB of virtual memory for each
of the executions. 
The experiments for different random seeds run in parallel (five seeds
at a time), with two unique cores available for every seed for
overheating reasons.
To take advantage of the decomposition scheme, \ac{DLNS} experiments
use eight threads (four physical cores) with three experiments (three
seeds at a time) running in parallel.
The rest of the algorithms run as sequential programs.
 For the larger benchmarks (Table~\ref{tab:benchmarks_large}), the
 available virtual memory for each of the executions is 64GB.
 The experiments for different random seeds run sequentially and the
 \ac{DLNS} experiments use eight threads.

\paragraph{Algorithm Configuration.}
The experiments focus on speed optimization and aim to generate 200
variants within a timeout.
Parameter $h$ in Algorithms~\ref{alg:search} and \ref{alg:decomp} is
set to one because even small distance between variants is able to
break gadgets (see Figure~\ref{fig:mips_example}).
\ac{LNS} uses restart-based search with a limit of 1000 failures and a
relax rate of 60\%.
The \textit{relax rate} is the probability that \ac{LNS} destroys a
variable at every restart, which affects the distance between two
subsequent solutions.
The relax rate is selected empirically based on preliminary experiments
(Appendix~\ref{sec:appa}).
Note that in our previous paper \shortcite{tsoupidi2020constraint},
the best relax rate on a different benchmark set was found to be 70\%.
This suggests that the optimal relax rate depends on the properties of
the program under compilation, where the number of instructions
appears to be a significant factor.
\ac{DLNS} uses the same parameters as \ac{LNS} for the
\textit{local} problems, which consist of the individual basic
blocks, and a relax rate of 50\% for the global problem.

\begin{table}
  \centering
   \caption{\label{tab:mediabench}Benchmarks functions - 10 to 100 \ac{MIR} instructions}
  \setlength\tabcolsep{2pt}
    \begin{tabular}{|l|l|l|r|r|}
\hline
{ID}&{application}&{function name}&{\# blocks}&{\# instructions}\\
\hline
b1&rasta&FR2TR&4&19
\\
b2&mesa&glColor3ubv&1&20
\\
b3&mesa&glTexCoord1dv&1&21
\\
b4&g721&ulaw2alaw&4&22
\\
b5&jpeg&start\_pass\_main&5&26
\\
b6&mesa&glTexCoord4sv&1&27
\\
b7&mesa&glEvalCoord2d&5&47
\\
b8&mesa&glTexGendv&5&58
\\
b9&rasta&open\_out&8&58
\\
b10&jpeg&quantize3\_ord\_dither&7&71
\\
b11&mpeg2&pbm\_getint&12&86
\\
b12&mesa&gl\_save\_PolygonMode&11&89
\\
b13&ghostscript&gx\_concretize\_DeviceCMYK&13&93
\\
b14&mesa&gl\_save\_MapGrid1f&11&96
\\
\hline
\end{tabular}

\end{table}

\begin{table}
  \centering
   \setlength\tabcolsep{2pt}
   \caption{\label{tab:benchmarks_large}{Benchmarks functions - 100 to 1000 \ac{MIR} instructions}}
    \begin{tabular}{|l|l|l|r|r|}
\hline
{ID}&{application}&{function name}&{\#blocks}&{\#instructions}\\
\hline
b15&mesa&gl\_xform\_normals\_3fv&10&107
\\
b16&jpeg&start\_pass\_1\_quant&34&215
\\
b17&mesa&apply\_stencil\_op\_to\_span&65&267
\\
b18&mesa&antialiased\_rgba\_points&39&366
\\
b19&mesa&gl\_depth\_test\_span\_generic&102&403
\\
b20&mesa&general\_textured\_triangle&40&890
\\
\hline
\end{tabular}

\end{table}

\subsection{RQ1. Scalability and Diversification Effectiveness of \ac{LNS} and \ac{DLNS}}
\label{ssec:evallns}

This section evaluates the diversification effectiveness and scalability of
\ac{LNS} and \ac{DLNS}
compared to incremental \textsc{MaxDiverse$k$Set} and \ac{RS}.  
Here, effectiveness is the ability to maximize the difference between the different variants generated by a given algorithm.
Scalability is related to the number of variants generated within a fixed time budget
and the total time required to generate the maximum number of variants.
This experiment uses \ac{HD} as the distance measure because \ac{HD} is a
general-purpose distance that may be valuable for different applications.

We measure the diversification effectiveness of these methods based on
the relative pairwise distance of the solutions.  Given a set of
solutions $S$ and a distance measure $\delta$, the pairwise distance
$d$ of the variants in $S$ is:

\begin{equation}
  d(\delta, S) = \left. \sum_{i=0}^{|S|}
  \sum_{j>i}^{|S|} \delta(s_i, s_j) \, / \, {\binom{|S|}{2}}\right..
  \label{eq:pairwise}
\end{equation}

The \emph{larger} this distance, the more diverse the solutions are,
and thus, diversification is more effective.
%
Tables~\ref{tab:dist_max_rs_lns} and \ref{tab:dist_max_rs_lns_large} shows the pairwise distance $d$
and diversification time $t$ (in seconds) for each benchmark and method. Each experiment uses a time budget of 20 minutes and an optimality gap of $p=10\%$.
The best values of $d$ (larger) and $t$ (lower) are marked in \textbf{bold} for the completed
experiments, whereas incomplete experiments are highlighted in \textit{italic}
and their number of variants in parenthesis.
A complete experiment is an experiment, where the algorithm was able to generate
the maximum number of 200 variants within the time limit for at least five of the random seeds.
The values of $d$ and $t$ correspond to the results for these random seeds.

\begin{small}
  \setlength\tabcolsep{3pt}
  \begin{longtable}{|l|c|c|c|c|c|c|c|c|}
\caption{\label{tab:dist_max_rs_lns} Distance and Scalability of \ac{LNS} and \ac{DLNS} against \ac{RS} and \textsc{MaxDiverse$k$Set} - 10 to 100 \ac{MIR} instructions }\\
\hline
\multirow{2}{*}{ID}&\multicolumn{2}{c|}{\textsc{MaxDiverse$k$Set}}&\multicolumn{2}{c|}{{RS}}&\multicolumn{2}{c|}{LNS (0.6)}&\multicolumn{2}{c|}{DLNS (0.6)}\\
\cline{2-9}
&$d$&$t(s)$&$d$&$t(s)$&$d$&$t(s)$&$d$&$t(s)$\\
\hline
b1&\textit{36.4$\pm$7.7} & - (2)&4.1$\pm$0.3 & \textbf{0.1$\pm$0.0}&\textbf{26.6$\pm$6.6} & 2.4$\pm$0.9&12.0$\pm$1.6 & 9.4$\pm$5.8
\\
b2&\textit{18.7$\pm$0.2} & - (4)&5.7$\pm$0.1 & \textbf{0.2$\pm$0.0}&\textbf{13.2$\pm$0.6} & 1.7$\pm$0.3&9.7$\pm$1.1 & 9.4$\pm$2.0
\\
b3&\textit{19.3$\pm$1.2} & - (3)&5.1$\pm$0.1 & \textbf{0.2$\pm$0.0}&\textbf{14.8$\pm$1.1} & 1.4$\pm$0.3&9.8$\pm$1.9 & 5.8$\pm$1.2
\\
b4&\textit{22.4$\pm$0.0} & - (27)&5.3$\pm$0.0 & \textbf{0.2$\pm$0.0}&\textbf{15.4$\pm$1.4} & 1.1$\pm$0.2&11.8$\pm$1.9 & 11.7$\pm$9.2
\\
b5&\textit{35.0$\pm$0.7} & - (2)&5.3$\pm$0.0 & \textbf{0.2$\pm$0.0}&\textbf{22.8$\pm$2.3} & 2.9$\pm$0.3&13.1$\pm$1.6 & 5.7$\pm$0.8
\\
b6&\textit{28.0$\pm$0.0} & - (2)&4.5$\pm$0.0 & \textbf{0.4$\pm$0.0}&\textbf{23.5$\pm$0.8} & 13.8$\pm$2.2&22.0$\pm$1.0 & 51.6$\pm$12.2
\\
b7&- & -&4.9$\pm$0.1 & \textbf{0.4$\pm$0.0}&\textbf{45.2$\pm$2.4} & 7.3$\pm$1.1&19.9$\pm$5.0 & 4.3$\pm$0.8
\\
b8&- & -&4.3$\pm$0.1 & \textbf{0.5$\pm$0.0}&\textbf{57.4$\pm$3.0} & 11.1$\pm$1.4&25.6$\pm$5.6 & 4.6$\pm$0.7
\\
b9&- & -&3.0$\pm$0.0 & \textbf{0.7$\pm$0.0}&\textbf{64.0$\pm$7.2} & 15.6$\pm$5.2&28.1$\pm$6.7 & 6.1$\pm$2.1
\\
b10&- & -&\textit{1.0$\pm$0.0} & - (3)&\textbf{160.9$\pm$16.0} & 332.1$\pm$88.8&30.4$\pm$14.3 & \textbf{7.6$\pm$0.9}
\\
b11&- & -&1.9$\pm$0.0 & \textbf{7.6$\pm$0.1}&\textbf{155.9$\pm$4.4} & 110.0$\pm$27.1&48.9$\pm$13.6 & 7.7$\pm$1.3
\\
b12&- & -&1.7$\pm$0.0 & \textbf{4.5$\pm$0.7}&\textbf{127.4$\pm$3.7} & 361.3$\pm$77.3&32.2$\pm$15.1 & 6.0$\pm$0.4
\\
b13&- & -&1.9$\pm$0.0 & \textbf{3.0$\pm$0.0}&\textbf{103.7$\pm$5.4} & 94.6$\pm$39.7&46.7$\pm$9.8 & 15.5$\pm$21.9
\\
b14&- & -&1.2$\pm$0.1 & \textbf{6.0$\pm$0.1}&\textbf{139.3$\pm$2.9} & 865.4$\pm$99.4&39.3$\pm$20.1 & 7.0$\pm$0.5
\\
\hline
\end{longtable}

\end{small}

\begin{small}
  \setlength\tabcolsep{3pt}
  \begin{longtable}{|l|c|c|c|c|c|c|c|c|}
\caption{\label{tab:dist_max_rs_lns_large} Distance and Scalability of \ac{LNS} and \ac{DLNS} against \ac{RS} and \textsc{MaxDiverse$k$Set} - 100 to 1000 \ac{MIR} instructions}\\
\hline
\multirow{2}{*}{ID}&\multicolumn{2}{c|}{\textsc{MaxDiverse$k$Set}}&\multicolumn{2}{c|}{{RS}}&\multicolumn{2}{c|}{LNS (0.6)}&\multicolumn{2}{c|}{DLNS (0.6)}\\
\cline{2-9}
&$d$&$t(s)$&$d$&$t(s)$&$d$&$t(s)$&$d$&$t(s)$\\
\hline
b15&- & -&\textit{1.0$\pm$0.0} & - (7)&\textit{278.5$\pm$4.2} & - (159)&\textbf{30.6$\pm$25.5} & \textbf{103.3$\pm$51.2}
\\
b16&- & -&- & -&- & -&\textbf{73.1$\pm$41.0} & \textbf{57.3$\pm$14.7}
\\
b17&- & -&2.7$\pm$0.2 & 318.8$\pm$0.2&\textit{375.4$\pm$13.4} & - (27)&\textbf{147.9$\pm$37.1} & \textbf{92.0$\pm$33.7}
\\
b18&- & -&- & -&- & -&\textbf{167.5$\pm$169.8} & \textbf{287.2$\pm$4.0}
\\
b19&- & -&1.0$\pm$0.0 & 2902.8$\pm$1.6&- & -&\textbf{222.8$\pm$48.6} & \textbf{139.3$\pm$22.8}
\\\hline
b20&\multicolumn{8}{c|}{Unison and DivCon cannot handle this function.}
\\
\hline
\end{longtable}

\end{small}

\paragraph{Scalability.}
The scalability results ($t$) show that only \ac{DLNS} is
scalable to large benchmarks, i.e.\ it is able to generate the
maximum of 200 variants for all benchmarks except for \textit{b20}.
Benchmark \textit{b20} contains a large number of \ac{MIR}
instructions and a small number of basic blocks (see
Table~\ref{tab:benchmarks_large}) and thus, exceeds Unison's solving
capability~\cite{lozano_combinatorial_2019}.
\ac{RS} and \ac{LNS} are scalable for the majority of the
benchmarks between 10 and 100 lines of \ac{MIR} instructions
(Table~\ref{tab:dist_max_rs_lns}).
In both benchmark sets, \textsc{MaxDiverse$k$Set} scales poorly, it
cannot generate 200 variants for any benchmark.
\textsc{MaxDiverse$k$Set} is able to find a small number of variants
for \textit{b1}-\textit{b6}.
However, it is not able to find any variant for the rest of the
benchmarks.
The first six benchmarks are small functions with less that 30 \ac{MIR}
instructions, whereas the rest of the benchmarks are larger and
consist or more than 47 instructions (see Table~\ref{tab:mediabench}).

\ac{LNS} is slower than \ac{RS} and \ac{DLNS}, requiring up to 855
seconds or approximately 14.25 minutes for diversifying \textit{b14}.
Similar to \textsc{MaxDiverse$k$Set}, the number of instructions
appears to be the main factor that determines the scalability of
\ac{LNS}.
For the large benchmarks of Table~\ref{tab:dist_max_rs_lns},
\textit{b10}-\textit{b12}, and \textit{b14}, the diversification time is
larger than one minute, whereas for smaller benchmarks
\textit{b1}-\textit{b9}, which have less than 60
\ac{MIR} instructions, the diversification time is less than one
minute.
For the largest benchmarks (Table~\ref{tab:dist_max_rs_lns_large}),
\ac{LNS} is able to generate 159 variants for \textit{b15} in around
4.63 minutes, but is not able to scale for larger benchmarks.

\ac{DLNS} is generally slower than \ac{RS} for the
benchmarks of Table~\ref{tab:dist_max_rs_lns}, but is able to scale
to larger benchmarks, as seen in Table~\ref{tab:dist_max_rs_lns_large}, where \ac{RS}
manages to generate 200 variants only for \textit{b17} and \textit{b19}.
We can see that \ac{DLNS} has similar performance regardless of the
benchmark size, with a general increase in the diversification time
for larger benchmarks (Table~\ref{tab:dist_max_rs_lns_large}).
This increase depends on the number of threads (eight) that is smaller
than the number of basic blocks.
For small benchmarks with basic blocks that contain few instructions,
decomposition is not advantageous because it does not reduce the
search space significantly.
Instead, \ac{DLNS} introduces an overhead when some versions of the
local solutions cannot be combined into a solution.
Among the commonly scalable benchmarks, the advantage of \ac{DLNS}
compared to \ac{RS} is clear in medium and large benchmarks,
\textit{b10}-\textit{b14}, \textit{b17}, and
\textit{b19}, where \ac{DLNS} is able generate a large number of
variants.
At the same time, \ac{DLNS} demonstrates a large variation in the
solutions with the different seeds.
This is due to the decomposition scheme of Algorithm~\ref{alg:decomp}.
That is, depending on the random seed, the algorithm might need to
restart the global problem just once or multiple times.

Overall, for small benchmarks, i.e.\ less than 60 \ac{MIR}
instructions, \ac{RS}, \ac{LNS}, and \ac{DLNS} are all able to
generate program variants efficiently (less than 16 seconds), whereas
for larger benchmarks, only \ac{DLNS} is able to generate a large
number of variants efficiently.

\paragraph{Diversity.}
The diversity results ($d$) show that \ac{LNS} is more effective at
diversifying than \ac{RS} and \ac{DLNS}.  The improvement of \ac{LNS}
over \ac{RS} ranges from 1.3x (for \textit{b2}) to 115x (for
\textit{b14}), whereas the improvement of \ac{LNS} over \ac{DLNS} is
smaller and ranges from $7\%$ (for \textit{b6}) to $429\%$ or 4x (for
\textit{b10}). 
\ac{DLNS} is clearly less effective at generating highly diverse
variants than \ac{LNS}, but more effective than \ac{RS}.  In
particular, the improvement of \ac{DLNS} over \ac{RS} ranges from
$70\%$ (for \textit{b1}) to 222x (for \textit{b19}).
The difference between \ac{LNS} and \ac{DLNS} in
generating diverse solutions 
is due to the ability of the former to consider the problem as
a whole, enabling more fine-grained solutions.

\textsc{MaxDiverse$k$Set} is not able to generate
200 variants for any of the benchmarks but may give an
indication of an upper bound for diversification of the smaller
benchmarks.
That is, although \textsc{MaxDiverse$k$Set}
is not exact, i.e.\ it maximizes the pairwise distance iteratively, we expect that
\ac{LNS}, \ac{DLNS},  and \ac{RS}  are not able to achieve
higher pairwise diversity than \textsc{MaxDiverse$k$Set}.
However, a direct comparison is not possible because \textsc{MaxDiverse$k$Set}
is not able to generate 200 variants for any of the benchmarks.

\paragraph{Conclusion.}
In summary, \ac{LNS} and \ac{DLNS} provide two attractive solutions
for diversifying code: \ac{LNS} is significantly and consistently more
effective at diversifying code than both \ac{RS} and \ac{DLNS}, but
does not scale efficiently for large benchmarks, whereas \ac{DLNS} is
more effective than both \ac{LNS} and \ac{RS} at generating variants
for large problems, and is still able to improve significantly the
diversity over \ac{RS}.

\subsection{RQ2. Scalability of LNS with Different Distance Measures}
\label{ssec:scale_dist}

In this section, we compare the distance measures introduced in
Section~\ref{ssec:distances} with regards to their
ability to steer the search towards diverse program variants  within a maximum time budget. 
Based on the results of RQ1, we focus on the   \ac{LNS} search algorithm, and run it with each distance metric.
For the
  problem-specific distance measure, \ac{GD}, we compare two
  configurations, i) $n_r=0$ and $n_c = 2$ and ii) $n_r=0$ and $n_c =
  8$. The two parameters control the number of instructions preceding a
  branch that differ among different solutions.
  The smaller these parameters are, the higher the chance of breaking a larger number of gadgets,
given that all gadgets end with a branch instruction. 

\begin{figure*}[t]
  \centering
\begin{small}
  \setlength\tabcolsep{2pt}
    \begin{longtable}{|l|c|c|c|c|c|c|c|c|}
\caption{\label{tab:distances}{Scalability of $\delta_{HD}$, $\delta_{LD}$, $\delta_{GD}^{0,2}$, and $\delta_{GD}^{0,8}$}}\\
\hline
\multirow{2}{*}{ID}&\multicolumn{2}{c|}{$\delta_{HD}$}&\multicolumn{2}{c|}{$\delta_{LD}$}&\multicolumn{2}{c|}{$\delta_{GD}^{0,2}$}&\multicolumn{2}{c|}{$\delta_{GD}^{0,8}$}\\
\cline{2-9}
&$t(s)$&num&$t(s)$&num&$t(s)$&num&$t(s)$&num\\
\hline
b1&\textbf{2.7$\pm$0.9} & 200 &- & 37&6.9$\pm$7.1 & 200 &\textbf{2.9$\pm$1.0} & 200 
\\
b2&\textbf{1.8$\pm$0.4} & 200 &- & 41&- & 75&5.8$\pm$2.6 & 200 
\\
b3&\textbf{1.6$\pm$0.2} & 200 &- & 44&- & 121&4.5$\pm$3.4 & 200 
\\
b4&\textbf{1.3$\pm$0.2} & 200 &- & 38&2.5$\pm$0.9 & 200 &\textbf{1.4$\pm$0.4} & 200 
\\
b5&\textbf{3.6$\pm$0.3} & 200 &- & 27&- & 12&112.4$\pm$126.8 & 200 
\\
b6&\textbf{14.1$\pm$2.3} & 200 &- & 15&172.1$\pm$179.4 & 200 &17.6$\pm$3.3 & 200 
\\
b7&\textbf{7.9$\pm$1.3} & 200 &- & 12&181.5$\pm$183.4 & 200 &19.6$\pm$4.0 & 200 
\\
b8&\textbf{12.1$\pm$1.5} & 200 &- & 8&73.1$\pm$22.2 & 200 &32.1$\pm$6.6 & 200 
\\
b9&\textbf{17.0$\pm$4.6} & 200 &- & 5&- & 56&217.5$\pm$158.6 & 200 
\\
b10&348.6$\pm$90.7 & 200 &- & -&359.8$\pm$59.4 & 200 &\textbf{319.3$\pm$81.8} & 200 
\\
b11&\textbf{121.1$\pm$29.0} & 200 &- & -&- & 77&445.1$\pm$64.6 & 200 
\\
b12&\textbf{377.9$\pm$76.7} & 200 &- & -&- & 105&- & 60
\\
b13&\textbf{107.6$\pm$44.1} & 200 &- & -&377.7$\pm$158.4 & 200 &208.7$\pm$110.6 & 200 
\\
b14&- & 152&- & -&- & 55&- & 36
\\
\hline
\end{longtable}

\end{small}
\end{figure*}

Table~\ref{tab:distances} presents the results of the distance
evaluation, where the time limit is 10 minutes and the optimality gap
$p=10\%$.  For each distance measure ($\delta_{HD}$,
$\delta_{LD}$, $\delta_{GD}^{0,2}$, and $\delta_{GD}^{0,8}$), the table
shows the diversification time $t$, in seconds (or ``-'' if the
algorithm is not able to generate 200 variants) and the number of
generated variants $num$ within the time limit. The value of $num$
shows the maximum number of variants that at least five (out of 15) of the random seeds
generate.

The results show that when \ac{DivCon} uses \ac{LNS} with Hamming
Distance, $\delta_{HD}$, it generates 200 variants for all benchmarks
except \emph{b12}, where it generates 157 variants.  The
diversification time with $\delta_{HD}$ ranges from one second for
\textit{b4} to approximately six minutes for \textit{b12}.  On the
other hand, \ac{DivCon} using \acf{LD}, $\delta_{LD}$, is not able to
generate 200 variants for any of the benchmarks within the time limit.
The scalability issues of $\delta_{LD}$ are due to the quadratic
complexity of its implementation~\cite{wagner1974string}, whereas
Hamming Distance can be implemented linearly.  \ac{DivCon} using the
first configuration of \acf{GD}, $\delta_{GD}^{0,2}$, generates the
maximum number of variants for seven benchmarks,
i.e. \textit{b1},\textit{b4}, \textit{b6}-\textit{b8}, 
\textit{b10}, and \textit{b13}.
Distance $\delta_{GD}^{0,2}$ uses small values for parameters $n_r=0$
and $n_c=2$, which leads to a reduced number of solutions (see
Section~\ref{ssec:distances}).
This has a negative effect on the scalability, resulting in low
variant generation for the rest of the benchmarks.
Using the second configuration of \ac{GD}, distance
$\delta_{GD}^{0,8}$, with $n_r=0$ and $n_c=8$, \ac{DivCon} generates
the maximum number of variants for all benchmarks except \textit{b12}
and \textit{b14}.
The time to generate the variants with $\delta_{GD}^{0,8}$ is larger
than with $\delta_{HD}$.
With this gadget-targeting metric, \ac{DivCon} takes up to seven
minutes for generating 200 variants for \textit{b11}.

\paragraph{Conclusion.} \ac{DivCon} using  \ac{LNS} and the
   $\delta_{GD}^{0,8}$ or $\delta_{HD}$ distance can generate a large
number of diverse program variants for most of the benchmark
functions. Scalability, given the maximum number of variants, comes
with slightly longer diversification time for $\delta_{GD}^{0,8}$ than
with $\delta_{HD}$.
In Section~\ref{ssec:gadget_dist}, we evaluate the distance measures
with regards to security.

\subsection{RQ3. \ac{JOP} Attacks Mitigation: Effectiveness of LNS and DLNS}
\label{ssec:seval_algo}

Software Diversity has various applications in security, including
mitigating code-reuse attacks.
To measure the level of mitigation that \ac{DivCon} achieves, we assess the
\ac{JOP}
gadget survival rate $srate(s_i,s_j)$ between two variants $s_i,s_j \in S$,
where $S$ is the set of generated variants.  This metric determines how many of
the gadgets of variant $s_i$ appear at the same position on the other variant
$s_j$, that is $srate(s_i,s_j) = |gad(s_i) - gad(s_j)| \, / \, |gad(s_i)|$,
where $gad(s_i)$ are the gadgets in solution $s_i$.  The procedure for computing
$srate(s_i,s_j)$ is as follows: 1)  find
the set of gadgets $gad(s_i)$ in solution $s_i$, and 2) for every $g\in
gad(s_i)$, check whether there exists a gadget identical to $g$ at the same
address of $s_j$.
For step 1, we use the state-of-the-art tool,
ROPgadget~\cite{ROPGadget2020}, to automatically find the gadgets in
the \texttt{.text} section of the compiled code.
For step 2, the comparison is syntactic after removing all
\texttt{nop} instructions.
Syntactic comparison is scalable but may result in false
negatives.


This and the following sections evaluate the effectiveness of
\ac{DivCon} against code-reuse attacks.  To achieve this, all
experiments compare the distribution of $srate$ for all pairs of
generated solutions.  Due to its skewness, the distribution of $srate$
is represented as a histogram with four buckets (0\%, (0\%, 10\%],
  (10\%,40\%], and (40\%, 100\%]) rather than summarized using common
      statistical measures.  Here, the best is an $srate(s_i,s_j)$ of
      0\%, which means that $s_j$ does not contain any gadgets that
      exist in $s_i$, whereas an $srate(s_i,s_j)$ in range
      (40\%,100\%] means that $s_j$ shares more than 40\% of the
        gadgets of $s_i$.
      
To evaluate the gadget diversification efficiency, we compare the $srate$
for all permutations
of pairs in $S$
for \ac{LNS} and \ac{DLNS} with \ac{RS} as a baseline.
Low $srate$ corresponds to higher mitigation effectiveness
because code-reuse attacks based on gadgets in one variant
have lower chances of locating the same gadgets in the other variants
(see Figure~\ref{fig:mips_example}).
%
%
Tables~\ref{tbl:gadgets_methods} and~\ref{tbl:gadgets_methods_large}
summarize the gadget survival distribution
for all benchmarks for algorithms \ac{RS}, \ac{LNS}, and \ac{DLNS}.
We use 10\% as the optimality gap and \ac{HD} because, as we saw in
RQ2, \ac{DivCon} using \ac{HD} is the most scalable diversification
configuration.
The values in \textbf{bold} correspond to the most frequent value(s) of the histogram.
The time limit for this experiment is 20 minutes.
Column \textit{num} shows the average of the generated number of variants for all
random seeds.

First, we notice that for the smaller benchmarks, \textit{b2} to
\textit{b3}, and \textit{b6}, all algorithms are able to generate
variant pairs that share no gadgets, i.e.\ the most frequent values
are in the first bucket (column \textit{=0}). 
\ac{RS} generates diverse variants that share a small number of
gadgets for \textit{b2-b4}, \textit{b6}, and \textit{b10} (only three
variants).
For the other benchmarks, the most common values are in
the second (\textit{b11}), or the third (\textit{b5},
\textit{b7}-\textit{b9}, \textit{b12}-\textit{b14}, \textit{b17},
and \textit{b19}) bucket, which provides poor mitigation effectiveness
against \ac{JOP} attacks.
The poor effectiveness of \ac{RS} against code-reuse attacks can be
correlated with the poor diversity effectiveness of the method (see
Section~\ref{ssec:evallns}).

\ac{LNS} generates diverse variants that do not share any gadgets
(belong to the first bucket) for all benchmarks except \textit{b5}.
Benchmark \textit{b5} has different behavior because it has a highly
constrained register allocation due to specific constraints imposed by
the calling conventions.

Finally, \ac{DLNS} has similar performance to \ac{RS} for medium size
benchmarks (Table~\ref{tbl:gadgets_methods}),  but worse performance
for large benchmarks (Table~\ref{tbl:gadgets_methods_large}).
In particular, only five benchmarks \textit{b1}-\textit{b4} and
\textit{b6} are mostly in the first bucket.
Although \ac{DLNS} has relatively high pairwise distance (see
Table~\ref{tab:dist_max_rs_lns}), its effectiveness against code-reuse
attacks is low.
This is because in many small programs with a large number of basic
blocks, the number of registers that are shared among different basic
blocks (and thus assigned by the \textit{global} problem, see
Algorithm~\ref{alg:decomp}) is high, resulting in low diversity of the
register allocation among variants.

\paragraph{Conclusion.}
The \ac{LNS} diversification algorithm is significantly more effective
than both \ac{DLNS} and \ac{RS} at generating binary variants that
share a minimal number of JOP gadgets.

\begin{table}
  \setlength\tabcolsep{2pt}
  \begin{filecontents*}{tables/hist_methods_output0.6hamming10.csv}
,\tiny{$=$0},\tiny{$\le$10},\tiny{$\le$40},\tiny{$\le$100},\tiny{num},\tiny{$=$0},\tiny{$\le$10},\tiny{$\le$40},\tiny{$\le$100},\tiny{num},\tiny{$=$0},\tiny{$\le$10},\tiny{$\le$40},\tiny{$\le$100},\tiny{num}
b1,\textbf{34},13,19,\textbf{33},200,\textbf{85},12,2,-,200,\textbf{50},24,25,1,200
b2,\textbf{86},7,5,1,200,\textbf{88},1,11,1,200,\textbf{83},1,11,5,200
b3,\textbf{84},11,5,-,200,\textbf{90},3,6,-,200,\textbf{88},4,7,1,200
b4,\textbf{92},7,1,-,200,\textbf{95},4,1,-,200,\textbf{52},38,8,3,200
b5,2,5,\textbf{48},\textbf{45},200,14,14,\textbf{51},21,200,-,13,\textbf{44},\textbf{43},200
b6,\textbf{74},18,8,-,200,\textbf{92},3,5,-,200,\textbf{92},3,4,1,200
b7,-,26,\textbf{72},2,200,\textbf{87},11,2,-,200,7,23,\textbf{52},18,200
b8,-,36,\textbf{63},1,200,\textbf{88},10,2,-,200,7,22,\textbf{48},23,200
b9,-,10,\textbf{83},8,200,\textbf{57},24,18,1,200,3,11,\textbf{49},36,200
b10,\textbf{68},2,11,19,3,\textbf{98},-,1,1,200,22,1,6,\textbf{71},200
b11,-,\textbf{72},28,-,200,\textbf{73},23,3,-,200,4,5,41,\textbf{51},200
b12,-,-,\textbf{99},1,200,\textbf{80},18,2,-,200,1,8,\textbf{59},32,187
b13,26,9,\textbf{35},\textbf{30},200,\textbf{92},4,3,-,200,31,11,19,\textbf{39},149
b14,-,-,\textbf{98},2,200,\textbf{77},21,2,-,200,-,3,\textbf{61},36,179
\end{filecontents*}
\pgfplotstabletypeset[
    col sep=comma,
    string type,
    columns={0,1,2,3,4,5,6,7,8,9,10,11,12,13,14,15},
    columns/0/.style= {column type={|M{16pt}|}},
    columns/1/.style= {column type={M{16pt}}},
    columns/2/.style= {column type={M{14pt}}},
    columns/3/.style= {column type={M{14pt}}},
    columns/4/.style= {column type={M{17pt}}},
    columns/5/.style= {column type={|M{16pt}|}},
    columns/6/.style={column type={M{16pt}}},
    columns/7/.style={column type={M{14pt}}},
    columns/8/.style={column type={M{14pt}}},
    columns/9/.style={column type={M{17pt}}},
    columns/10/.style={column type={|M{16pt}|}},
    columns/11/.style={column type={M{16pt}}},
    columns/12/.style={column type={M{14pt}}},
    columns/13/.style={column type={M{14pt}}},
    columns/14/.style={column type={M{17pt}}},
    columns/15/.style={column type={|M{16pt}|}},
    header=false,
    every first row/.style={after row=\hline},
    every head row/.style={before row=   \caption{\label{tbl:gadgets_methods}
        Gadget survival rate for 10\% optimality gap with Hamming distance for \ac{RS},
        \ac{LNS}, and \ac{DLNS}  - 10 to 100 \ac{MIR} instructions}\\
      \hline\multirow{2}{*}{ID} & \multicolumn{5}{c|}{\small{RS}} & \multicolumn{5}{c|}{\small{LNS}} & \multicolumn{5}{c|}{\small{DLNS}}\\\cline{2-16},output empty row},
    every last row/.style={after row=\hline},
    font=\small,
  ]{tables/hist_methods_output0.6hamming10.csv}
\end{table}

  \begin{table}
    \setlength\tabcolsep{2pt}
    \caption{\label{tbl:gadgets_methods_large} Gadget survival rate for 10\% optimality gap with Hamming distance for \ac{RS}, \ac{LNS}, and \ac{DLNS} - 100 to 1000 \ac{MIR} instructions}
  \centering
  \begin{small}
      \begin{tabular}{|c|cccc|c|cccc|c|cccc|c|}
  \hline\multirow{2}{*}{ID} & \multicolumn{5}{c|}{\small{RS}} & \multicolumn{5}{c|}{\small{LNS}} & \multicolumn{5}{c|}{\small{DLNS}}\\ \cline{2-16}
&\scriptsize{$=$0}&\scriptsize{$\le$10}&\scriptsize{$\le$40}&\scriptsize{$\le$100}&\scriptsize{num}&\scriptsize{$=$0}&\scriptsize{$\le$10}&\scriptsize{$\le$40}&\scriptsize{$\le$100}&\scriptsize{num}&\scriptsize{$=$0}&\scriptsize{$\le$10}&\scriptsize{$\le$40}&\scriptsize{$\le$100}&\scriptsize{num}\\\hline
b15&\textbf{98}&-&2&-&7&\textbf{99}&1&-&-&118&\textbf{43}&2&7&\textbf{47}&188\\
b16&-&-&-&-&-&\textbf{98}&2&-&-&71&-&1&33&\textbf{66}&200\\
b17&30&13&\textbf{47}&10&200&\textbf{87}&6&6&1&42&15&10&35&\textbf{40}&173\\
b18&-&-&-&-&-&-&-&-&-&-&-&-&2&\textbf{98}&187\\
b19&18&27&\textbf{52}&3&200&-&-&-&-&-&1&-&40&\textbf{60}&200\\\hline
b20&\multicolumn{15}{c|}{Unison and DivCon cannot handle this function.}\\\hline 
\end{tabular}

  \end{small}
  \end{table}

\subsection{RQ4. \ac{JOP} Attacks Mitigation: Effectiveness of Different Distance Measures}
\label{ssec:gadget_dist}

Section~\ref{ssec:scale_dist} shows that 
\acf{HD}, $\delta_{HD}$, is the most scalable distance measure followed closely by
the second configuration of \acf{GD}, $\delta_{GD}^{0,8}$.
This section investigates the impact of the distance measure on the effectiveness of
\ac{DivCon} against \ac{JOP} attacks.

Table~\ref{tbl:gadget_dist} shows the gadget-replacement effectiveness
of \ac{DivCon} using distances: $\delta_{HD}$, $\delta_{LD}$, $\delta_{GD}^{0,2}$,
and $\delta_{GD}^{0,8}$. The time limit for this experiment is ten
minutes and the optimality gap is 10\%. This experiment uses \ac{LNS} as the diversification
algorithm because, as we have seen in Section~\ref{ssec:seval_algo},
\ac{LNS} is more effective against \ac{JOP} attacks than \ac{DLNS}.

The results for the Hamming Distance (\ac{HD}), $\delta_{HD}$, are in the
first column of the table.
For all benchmarks, except \textit{b5}, the highest values are under
the first subcolumn. This means that a large proportion of the variant pairs
do not share any gadgets, which is a strong indication of \ac{JOP} attack mitigation.
In particular, the most frequent values range from 57 to 98 percent.
Benchmark \textit{b5} has
weak diversification capability due to hard constraints in register allocation
(see  Section~\ref{ssec:seval_algo}).

The results for \acl{LD}, $\delta_{LD}$, appear in the second column
of the table.  Similar to \ac{HD}, almost all benchmarks,
where \ac{DivCon} generates at least two variants, have their
most common value in the first subcolumn except for \textit{b5}.
These values range from
51\% to 85\%, which corresponds to poorer gadget diversification effectiveness
than 
using $\delta_{HD}$.
As discussed in Section~\ref{ssec:scale_dist}, \ac{DivCon} using \acl{LD}
is not able to generate the maximum requested number of variants (200)
within the time limit of ten minutes for any of the benchmarks.

The third column of Table~\ref{tbl:gadget_dist} shows the results for \acf{GD} with parameters
$n_r=0$ and $n_c=2$. Parameter $n_r = 0$ enforces
diversity of the register allocation for the instructions
that are issued on the same cycle as the branch instruction. 
Similarly, parameter $n_c = 2$ enforces diversity for the
instruction schedule of the instructions
preceding the branch instruction by at most two cycles.
Distance $\delta_{GD}^{0,2}$ measures the sum of these two constraints
(and enforces it to be greater than $h=1$)
for all branch instructions of the
benchmark in question.
\ac{DivCon} with this distance measure has
very high effectiveness against JOP attacks, with the most frequent values ranging from 65 to 100 percent. 
However, using $\delta_{GD}^{0,2}$, \ac{DivCon}
is not able to generate a large number of variants for almost half of the
benchmarks.

The last distance measure, $\delta_{GD}^{0,8}$, differs from
$\delta_{GD}^{0,2}$ in that it allows diversifying the instruction
schedule for a larger number of instructions preceding the branch
instruction, i.e.\ $n_c=8$.
Here, the most common values range from 48 to 99 percent for different
benchmarks and the scalability is satisfiable with \ac{DivCon} being
able to generate the total number of requested variants for almost all
the benchmarks.
Using $\delta_{GD}^{0,8}$, \ac{DivCon} improves the gadget
diversification efficiency of the overall fastest distance measure,
$\delta_{HD}$, for all benchmarks except \textit{b3}, where the
difference is very small.
The largest improvement is for \textit{b9} and \textit{b5}. For
\textit{b9} the most frequent value is 57\% with $\delta_{HD}$ and
gets improved to 66\% with $\delta_{GD}^{0,8}$.
For \textit{b5} the majority of the variant pairs are under the third
bucket, which corresponds to the weak $(10\%-40\%]$-survival rate with
  $\delta_{HD}$ and under the first bucket (column \textit{=0}) with
  $\delta_{GD}^{0,8}$, which is a significant improvement.

\paragraph{Conclusion.}
Distances $\delta_{HD}$ and $\delta_{GD}^{0,8}$
are both appropriate distances for our application,
trading scalability with security effectiveness.
\ac{DivCon} using $\delta_{HD}$ has better scalability than using $\delta_{GD}^{0,8}$
(see Section~\ref{ssec:scale_dist}), whereas \ac{DivCon} using  $\delta_{GD}^{0,8}$
is more effective against code-reuse attacks compared to using $\delta_{HD}$.

\begin{table}[t]
  \setlength\tabcolsep{1pt}
 \begin{filecontents*}{tables/hist_metrics_output0.610.csv}
,\tiny{$=$0},\tiny{$\le$10},\tiny{$\le$40},\tiny{$\le$100},\tiny{num},\tiny{$=$0},\tiny{$\le$10},\tiny{$\le$40},\tiny{$\le$100},\tiny{num},\tiny{$=$0},\tiny{$\le$10},\tiny{$\le$40},\tiny{$\le$100},\tiny{num},\tiny{$=$0},\tiny{$\le$10},\tiny{$\le$40},\tiny{$\le$100},\tiny{num}
b1,\textbf{85},12,2,-,200,\textbf{52},30,10,8,29,\textbf{99},1,-,-,200,\textbf{92},7,1,-,200
b2,\textbf{88},1,11,1,200,\textbf{85},-,12,2,37,\textbf{94},4,2,-,67,\textbf{90},4,6,-,200
b3,\textbf{90},3,6,-,200,\textbf{85},5,9,1,41,\textbf{95},4,1,-,111,\textbf{89},6,5,-,200
b4,\textbf{95},4,1,-,200,\textbf{85},12,2,1,36,\textbf{99},1,-,-,200,\textbf{97},3,-,-,200
b5,14,14,\textbf{51},21,200,15,10,32,\textbf{42},25,\textbf{65},9,24,2,12,\textbf{48},28,20,3,178
b6,\textbf{92},3,5,-,200,\textbf{84},4,10,2,14,\textbf{96},3,1,-,187,\textbf{92},4,4,-,200
b7,\textbf{87},11,2,-,200,\textbf{54},28,16,2,10,\textbf{83},15,2,-,145,\textbf{87},12,1,-,200
b8,\textbf{88},10,2,-,200,\textbf{53},23,20,4,7,\textbf{87},12,1,-,188,\textbf{88},11,1,-,200
b9,\textbf{57},24,18,1,200,\textbf{51},11,21,17,4,\textbf{74},15,11,-,52,\textbf{66},24,10,-,167
b10,\textbf{98},-,1,1,200,-,-,-,-,-,\textbf{99},-,-,-,200,\textbf{99},-,1,1,200
b11,\textbf{73},23,3,-,200,-,-,-,-,-,\textbf{91},8,1,-,62,\textbf{79},20,2,-,198
b12,\textbf{80},18,2,-,200,-,-,-,-,-,\textbf{96},4,-,-,83,\textbf{87},12,1,-,48
b13,\textbf{92},4,3,-,200,-,-,-,-,-,\textbf{100},-,-,-,185,\textbf{97},1,2,-,200
b14,\textbf{77},21,2,-,141,-,-,-,-,-,\textbf{95},5,-,-,44,\textbf{85},14,1,-,31
\end{filecontents*}
\pgfplotstabletypeset[
   col sep=comma,
   string type,
   columns={0,1,2,3,4,5,6,7,8,9,10,11,12,13,14,15,16,17,18,19,20},
   columns/0/.style= {column type={|M{16pt}|}},
   columns/1/.style= {column type={M{16pt}}},
   columns/2/.style= {column type={M{14pt}}},
   columns/3/.style= {column type={M{14pt}}},
   columns/4/.style= {column type={M{17pt}}},
   columns/5/.style= {column type={|M{16pt}|}},
   columns/6/.style={column type={M{16pt}}},
   columns/7/.style={column type={M{14pt}}},
   columns/8/.style={column type={M{14pt}}},
   columns/9/.style={column type={M{17pt}}},
   columns/10/.style={column type={|M{16pt}|}},
   columns/11/.style={column type={M{16pt}}},
   columns/12/.style={column type={M{14pt}}},
   columns/13/.style={column type={M{14pt}}},
   columns/14/.style={column type={M{17pt}}},
   columns/15/.style={column type={|M{16pt}|}},
   columns/16/.style={column type={M{16pt}}},
   columns/17/.style={column type={M{14pt}}},
   columns/18/.style={column type={M{14pt}}},
   columns/19/.style={column type={M{17pt}}},
   columns/20/.style={column type={|M{16pt}|}},
   header=false,
   every first row/.style={after row=\hline},
   every head row/.style={before row=\caption{\label{tbl:gadget_dist}
       Gadget survival rate for 10\% optimality gap
    for the distances: $\delta_{HD}$, $\delta_{LD}$,
    $\delta_{GD}^{0,2}$, and $\delta_{GD}^{0,8}$}\\
     \hline\multirow{2}{*}{ID} & \multicolumn{5}{c|}{\small{$\delta_{HD}$}} & \multicolumn{5}{c|}{\small{$\delta_{LD}$}} & \multicolumn{5}{c|}{\small{$\delta_{GD}^{0,2}$}} & \multicolumn{5}{c|}{\small{$\delta_{GD}^{0,8}$}}\\\cline{2-21},output empty row},
   every last row/.style={after row=\hline},
   font=\small,
 ]{tables/hist_metrics_output0.610.csv}

\end{table}

\subsection{RQ5. \ac{JOP} Attacks Mitigation: Effectiveness for Different Optimality Gaps}
\label{ssec:seval_gaps}

This section investigates the trade-off between code quality and
diversity and evaluates the effectiveness of \ac{DivCon} against
code-reuse attacks.
Table~\ref{tbl:optimal} summarizes the gadget survival distribution
for all benchmarks and different values of the optimality gap (0\%,
5\%, 10\%, and 20\%).
Based on the results of RQ3, we select \ac{LNS} for this evaluation
because we have observed that \ac{DivCon} using \ac{LNS} is the most
effective at diversifying gadgets.
Similarly, in RQ4, we were able to identify that the gadget-specific
distance, $\delta_{GD}^{0,8}$, is the most effective among the two
scalable distance measures at diversifying gadgets.
The values in \textbf{bold} correspond to the mode(s) of the histogram
and the time limit for this experiment is ten minutes.

First, we notice that \ac{DivCon} with \ac{LNS} and $\delta_{GD}^{0,8}$
can generate some pairs of variants
that share no gadgets, even without relaxing the constraint of
optimality ($p = 0\%$).
In particular, for $p = 0\%$, all benchmarks except
\textit{b7} are dominated by a $0\%$ survival rate
and only \textit{b7} is dominated by a weak $(0\%-10\%]$-survival rate.
This indicates that
optimal code naturally includes software diversity that is good for
security.
For example, \ac{DivCon} generates on average 110 solutions for
benchmark \textit{b6}.
Comparing pairwise the gadgets for these solutions, we are able to
determine that 91 percent of the solution pairs do not share any
gadgets, whereas five percent of these pairs share up to 10\% of the
gadgets and four percent share between 10\% and 40\% of the gadgets.
Furthermore, we can see that for only two of the benchmarks
(\textit{b5} and \textit{b9}), \ac{DivCon} with \ac{LNS} and
$\delta_{GD}^{0,8}$ is unable to generate any variants, whereas for
three of the benchmarks (\textit{b1}, \textit{b3}, and \textit{b13})
it generates a large number of variants without quality loss.
Among the benchmarks that are dominated by the first bucket ($0\%$
gadget survival rate), the rates range from 52\% up to 100\%.
These results indicate that it is possible to achieve high
security-aware diversity without sacrificing code quality.

Second, the results show that the effectiveness of \ac{DivCon} at
diversifying gadgets can be further increased by relaxing the
constraint on code quality, with diminishing returns beyond $p =
10\%$.
Increasing the optimality gap to just $p = 5\%$ makes $0\%$ survival
rate (column \textit{=0}) the dominating bucket for all benchmarks
except \textit{b5}.  Benchmark \textit{b5} is subjected to hard
register allocation constraints, which reduces \ac{DivCon}'s
gadget diversification ability.
The rate of the variant pairs that do not share any variants ranges
from 59 percent for \textit{b9} to 99 percent for \textit{b10}.
Further increasing the gap to $10\%$ and $20\%$ increases
significantly the number of pairs that share no gadgets (column
\textit{=0}).
For example, with an optimality gap of $p = 10\%$, the dominating
bucket for all benchmarks corresponds to $0\%$ survival rate (column
\textit{=0}) and ranges from 48\% (\textit{b5}) to 99\% (\textit{b10}) of
the total solution pairs.
An optimality gap of $p = 20\%$ improves further the effectiveness of
\ac{DivCon}.
The improvement is substantial for benchmark \textit{b5}, where the
register allocation of this benchmark is highly constrained.  Larger
optimality gap allows the generation of more solutions that differ
with regards to the instructions schedule.
This leads to an improvement indicated by an increase in the rate of
the first bucket (column \textit{=0}) from 48\% for $p = 10\%$ to 66\%
for $p= 20\%$.

\begin{table}[t]
  \centering
  \setlength\tabcolsep{1pt}
  \caption{\label{tbl:optimal}
    Gadget survival rate for different optimality gap values of the Gadget Distance ($\delta_{GD}^{0,8}$) using LNS}
  \begin{tabular}{|c|cccc|c|cccc|c|cccc|c|cccc|c|}
\hline\multirow{2}{*}{ID} & \multicolumn{5}{c|}{\small{$p=0\%$}} & \multicolumn{5}{c|}{\small{$p = 5\%$}} & \multicolumn{5}{c|}{\small{$p = 10\%$}} & \multicolumn{5}{c|}{\small{$p = 20\%$}}\\\cline{2-21}
&\tiny{$=$0}&\tiny{$\le$10}&\tiny{$\le$40}&\tiny{$\le$100}&\tiny{num}&\tiny{$=$0}&\tiny{$\le$10}&\tiny{$\le$40}&\tiny{$\le$100}&\tiny{num}&\tiny{$=$0}&\tiny{$\le$10}&\tiny{$\le$40}&\tiny{$\le$100}&\tiny{num}&\tiny{$=$0}&\tiny{$\le$10}&\tiny{$\le$40}&\tiny{$\le$100}&\tiny{num}\\\hline
b1&\textbf{93}&3&4&-&200&\textbf{89}&9&2&-&200&\textbf{92}&7&1&-&200&\textbf{98}&1&1&-&200\\
b2&\textbf{93}&-&7&-&20&\textbf{90}&4&6&-&200&\textbf{90}&4&6&-&200&\textbf{90}&4&5&-&200\\
b3&\textbf{80}&13&6&1&149&\textbf{90}&5&5&-&200&\textbf{89}&6&5&-&200&\textbf{93}&3&4&-&200\\
b4&\textbf{98}&1&-&-&24&\textbf{97}&3&-&-&200&\textbf{97}&3&-&-&200&\textbf{98}&2&-&-&200\\
b5&-&-&-&-&-&10&13&\textbf{42}&35&29&\textbf{48}&28&20&3&178&\textbf{66}&18&14&2&200\\
b6&\textbf{91}&5&4&-&110&\textbf{92}&4&4&-&200&\textbf{92}&4&4&-&200&\textbf{94}&3&3&-&200\\
b7&38&\textbf{48}&14&-&82&\textbf{85}&14&1&-&200&\textbf{87}&12&1&-&200&\textbf{89}&10&1&-&200\\
b8&\textbf{60}&30&10&-&40&\textbf{89}&10&1&-&200&\textbf{88}&11&1&-&200&\textbf{90}&9&1&-&200\\
b9&-&-&-&-&-&\textbf{59}&28&13&-&171&\textbf{66}&24&10&-&167&\textbf{66}&23&11&-&167\\
b10&\textbf{75}&3&3&19&4&\textbf{99}&-&1&1&200&\textbf{99}&-&1&1&200&\textbf{99}&-&-&-&193\\
b11&\textbf{84}&14&2&-&87&\textbf{82}&17&2&-&190&\textbf{79}&20&2&-&198&\textbf{84}&14&1&-&199\\
b12&\textbf{82}&15&3&-&12&\textbf{90}&9&1&-&36&\textbf{87}&12&1&-&48&\textbf{90}&9&1&-&57\\
b13&\textbf{100}&-&-&-&175&\textbf{96}&1&2&-&200&\textbf{97}&1&2&-&200&\textbf{97}&1&1&-&200\\
b14&\textbf{52}&41&7&-&3&\textbf{88}&11&1&-&25&\textbf{85}&14&1&-&31&\textbf{91}&8&1&-&44\\\hline
\end{tabular}

\end{table}

Related approaches (discussed in Section~\ref{sec:rel}) report the
\emph{average} gadget elimination rate
across all pairs for different benchmark sets.
The zero-cost approach of \shortciteauthor{pappas_smashing_2012}~\citeyear{pappas_smashing_2012} achieves
an average gadget elimination rate between $74\%-83\%$ without code degradation, comparable to
\ac{DivCon}'s $93\%-100\%$ at $p=0\%$ (including only benchmarks for which
\ac{DivCon} generates variants).
\shortciteauthor{homescu_profile-guided_2013}~\citeyear{homescu_profile-guided_2013} propose a statistical approach that
reports an average $srate$ between $82\%-100\%$ with a code degradation of less
than $5\%$, comparable to \ac{DivCon}'s $62\%-100\%$ at $p=5\%$.
Both approaches report results on larger code bases that exhibit more
opportunities for diversification.
We expect that \ac{DivCon} would achieve higher overall survival rates
on these code bases compared to the benchmarks used in this paper as
we can see in case study of RQ6 (Section~\ref{ssec:casestudy}).

\paragraph{Conclusion.} Empirical evidence shows that \ac{DivCon} with the \ac{LNS} algorithm and distance measure 
$\delta_{GD}^{0,8}$ achieves high \ac{JOP} gadget diversification rate 
without sacrificing code quality. Increasing the optimality gap to just 5\% 
improves the effectiveness of \ac{DivCon} significantly, while 
further increase in the optimality gap does not have a similarly large effect
on gadget diversity.

\subsection{RQ6. Case Study: Effectiveness of DivCon at the application level}
\label{ssec:casestudy}
\begin{table}
  \centering
  \caption{\label{tab:benchmarks_g721}{G.721 functions}}
  \begin{small}
    \setlength\tabcolsep{2pt}
    \begin{tabular}{|l|l|l|l|r|r|r|r|}
\hline
ID  & app  & module &{function name}&{\#blocks}&{\#instructions}& LNS time (s) & DLNS time (s)\\
\hline
g1  & g721 & g711 &        ulaw2linear    &      1    &    14    & 0.4   $\pm$  0.0 & 7.8  $\pm$0.0\\
g2  & g721 & g711 &          alaw2ulaw    &      4    &    19    & 0.8   $\pm$  0.0 & 52.6 $\pm$0.0\\
g3  & g721 & g711 &          ulaw2alaw    &      4    &    22    & 1.3   $\pm$  0.0 & 34.8 $\pm$0.0\\
g4  & g721 & g711 &        alaw2linear    &      6    &    23    & 0.9   $\pm$  0.0 & 22.5 $\pm$0.0\\
g5  & g721 & g72x &        reconstruct    &      4    &    24    & 0.8   $\pm$  0.0 & 22.4 $\pm$0.0\\
g6  & g721 & g72x &          step\_size   &      7    &    27    & 3.2   $\pm$  0.0 & 7.1  $\pm$0.0\\
g7  & g721 & g72x &     predictor\_pole   &      1    &    28    & 2.4   $\pm$  0.0 & 15.8 $\pm$0.0\\
g8  & g721 & g72x &    g72x\_init\_state  &      1    &    29    & 1.1   $\pm$  0.0 & 3.1  $\pm$0.0\\
g9  & g721 & g711 &        linear2ulaw    &     11    &    54    & 6.0   $\pm$  0.0 & 9.1  $\pm$0.0\\
g10 & g721 & g711 &        linear2alaw    &     13    &    60    & 30.5  $\pm$  0.0 & 6.7  $\pm$0.0\\
g11 & g721 & g72x & tandem\_adjust\_ulaw  &      9    &    75    & 140.8 $\pm$  0.8 & 6.8  $\pm$0.0\\
g12 & g721 & g72x &     predictor\_zero   &      1    &    77    & 43.8  $\pm$  0.1 & 5.3  $\pm$0.0\\
g13 & g721 & g72x & tandem\_adjust\_alaw  &     13    &    89    & 182.1 $\pm$  0.9 & 8.1  $\pm$0.0\\
g14 & g721 & g72x &           quantize    &     23    &    99    & 246.2 $\pm$  0.2 & 17.9 $\pm$0.0\\
g15 & g721 & g721 &       g721\_encoder   &      7    &   135    & 214.7 $\pm$  0.4 & 11.0 $\pm$0.0\\
g16 & g721 & g721 &       g721\_decoder   &      7    &   135    & 323.3 $\pm$  6.3 & 10.7 $\pm$0.0\\
g17 & g721 & g72x &             update    &    105    &   523    &  -               & 128.0$\pm$1.1\\\hline
g18 & main & main &             main      &      9    &    40    & 7.3   $\pm$  0.0 & 7.8  $\pm$0.0\\
g19 & main & main &       pack\_output    &      3    &    23    & 0.8   $\pm$  0.0 & 6.5  $\pm$0.0\\\hline
g20 & stubs& stubs&        \_nmi\_handler &      2    &     1    & - (1)    & - (1) \\
g21 & stubs& stubs&       \_on\_bootstrap &      1    &     1    & - (1)    & - (1) \\
g22 & stubs& stubs&        \_on\_reset    &      1    &     1    & - (1)    & - (1) \\
\hline
\end{tabular}

  \end{small}
\end{table}


\ac{DivCon} operates at the function level.
In this section, we evaluate the effectiveness of \ac{DivCon}
against \ac{JOP} attacks for programs that consist of multiple functions.
To do that, we study an application from MediaBench I and
evaluate it using the JOP gadget survival rate as in RQ3, RQ4, and
RQ5.
To diversify a program, we diversify the functions that
comprise this program and then combine them randomly.
This approach results in up to $n^{f}$ different variants, where $n$ is
the number of variants per function and $f$ the number of functions
in the program.
If we also perform function permutation, the number of possible
program variants increases to $f! \cdot n^{f}$.

We apply these methods on G.721, an application of the
MediaBench I benchmark suite~\shortcite{Lee1997}.
This application is an implementation of the \ac{CCITT} G.711, G.721,
and G.723 voice compression algorithms.
We compile G.721 for the MIPS32-based Pic32MX
microcontroller\footnote{PIC32MX Microprocessor Family:
  \url{https://www.microchip.com/en-us/products/microcontrollers-and-microprocessors/32-bit-mcus/pic32-32-bit-mcus/pic32mx}}.

Table~\ref{tab:benchmarks_g721} shows 1) the functions that comprise
the G.721 application, 2) a custom \texttt{main} function\footnote{The
  main function is a simplified version of the encoding example that
  g721 provides.} that performs encoding, and 3) a number of required
system functions, \texttt{stubs}.
The columns show the number of basic blocks (\#blocks), the number of
\ac{MIR} instructions (\#instructions) and the diversification time in
seconds for generating 200 variants using LNS (LNS time (s)) and DLNS
(DLNS time (s)) after running the experiment five times with the same
random seed (seed = 42).
The \texttt{stubs} functions consist of two empty functions
(\texttt{on\_reset} and \texttt{on\_bootstrap}) and one function
(\texttt{nmi\_handler}) that contains one empty infinite loop.
These functions contain only one \ac{MIR} instruction each, and, therefore,
there are no variants within a 10\% optimality gap.
We diversify the rest of the functions using \ac{DivCon} with 0.5
relax rate, 10\% optimality gap, and a time limit of 20 minutes.
We run the experiment using the same random seed for \ac{DivCon} and
the function randomization.
For the cases that LNS manages to generate all variants (all but
\textit{g17}), we use the LNS-generated variants and for the rest of
the benchmarks (\textit{g17}), we use DLNS.
For compiling the application, we generate the textual assembly code
of the function variants using \ac{DivCon} and \texttt{llc}.
To compile and link the application, we use a Pic32MX microcontroller
toolchain\footnote{\url{https://github.com/is1200-example-projects/mcb32tools}}
that uses \texttt{gcc}.
To deactivate instruction reordering by \texttt{gcc}, \texttt{llc}
sets the \texttt{noreorder} directive.

For combining the functions in the final binaries, we use two
approaches, 1) \ac{NFS}, which generates the binary combining the
different function variants in the same order and 2) \ac{FS}, which
randomizes the function order at the linking time.

\begin{table}
  \centering
  \caption{\label{tbl:lib}{Gadget survival rate for 10\% optimality
      gap with the Hamming distance, $\delta_{HD}$, for the G.721
      application with function randomization at link level (FS) and
      without (NFS)}}
  \begin{small}
    \setlength\tabcolsep{2pt}
      \begin{tabular}{|l|cccccccc|c|cccccccc|c|}
    \hline
    \multirow{2}{*}{App}&\multicolumn{9}{c|}{\small{NFS}} & \multicolumn{9}{c|}{\small{FS}}\\
    \cline{2-19}
& \scriptsize{$=$0}& \scriptsize{$\le$0.5} & \scriptsize{$\le$1} & \scriptsize{$\le$2}& \scriptsize{$\le$5}& \scriptsize{$\le$10}& \scriptsize{$\le$40}& \scriptsize{$\le$100}& \scriptsize{num}& \scriptsize{$=$0}& \scriptsize{$\le$0.5} & \scriptsize{$\le$1} & \scriptsize{$\le$2}& \scriptsize{$\le$5}& \scriptsize{$\le$10}& \scriptsize{$\le$40}& \scriptsize{$\le$100}& \scriptsize{num}\\\hline
    G.721& \textbf{85}& 12& 1& 1& -& -& -& -& 200 &\textbf{98}& 2& -& -& -& -& -& -& 200 \\\hline
\end{tabular}

  \end{small}
\end{table}

  Table~\ref{tbl:lib} shows the results of the
  diversification of G.721 using the \ac{NFS} and \ac{FS} schemes
  after generating 200 variants of the G.721 application.
  The results show that combining the diversified variants without
  shuffling the functions at link time (\ac{NFS}) results in most of
  the variants, 85\% of the pairs, sharing no gadgets, while 12\%
  share between 0\% and 0.5\% of the gadgets.
  We calculate the average of gadget survival rate over the variant
  pairs as $0.068\pm 0.128\%$.
  Using function shuffling at link time (\ac{FS}) results in
  $0.008\pm 0.008\%$ average gadget survival rate, with 98\% of all variant pairs
  not sharing any gadget (first bucket).
  This shows that the fine-grained diversification of \ac{DivCon}
  using function shuffling improves further the result for \ac{NFS}.
\paragraph{Conclusion.}


In this case study, we show that with our method, we are able to
diversify whole programs and not just functions.
Additionally, we show that randomly combining the diversified
functions using \ac{DivCon} achieves the diversification and/or
relocation of JOP gadgets with an average of less than 0.1\% survival
rate in a multi-function program.
Function shuffling reduces further the gadget survival rate to
approximately 0.01\% survival rate, indicating that hardly any
variant pairs share gadgets.

\subsection{Discussion}

\paragraph{Advanced code-reuse attacks.}
\label{ssec:advrop}
Our attack model considers basic-ROP/JOP attacks.
However, in literature there exist more advanced
attacks, 
like
JIT-ROP~\cite{snow_just--time_2013}, where the attacker is able to read the code from the memory
and identify gadgets during the attack.
Static diversification of a binary is not effective against these
types of attacks.
Instead, some
approaches~\cite{chen_remix_2016,williams-king_shuffler_2016} use
re-randomization, a technique to re-randomize the binary by switching
between variants of the code at run time.
Using our approach, it is possible to perform re-randomization
of an application by switching between different function variants
that \ac{DivCon} generates.
%

\paragraph{Large Functions.}

Unison is not scalable to large functions for
MIPS~\cite{lozano_combinatorial_2019} and in this paper we have
evaluated \ac{DivCon} for functions up to 523 lines of LLVM \ac{MIR}
instructions.
However, there are functions that are larger than what Unison
supports.
In particular, in MediaBench I, approximately 7\% of the functions
contain more than 500 instructions.
For these cases, one may use other diversification schemes for just
these functions and \ac{DivCon} for the rest of the functions.
Another approach is to deactivate some of the transformations that
Unison and \ac{DivCon} perform for larger benchmarks or improve the
scalability of Unison~\cite{lozano_combinatorial_2019}.
We leave this as future work.

\section{Related Work}
\label{sec:rel}

State of the art software diversification
techniques 
apply randomized transformations 
at different stages of the software development. Only
a few exceptions use search-based techniques~\cite{larsen_sok_2014}.
This section focuses on quality-aware software diversification approaches.

\textit{Superdiversifier}~\shortcite{jacob_superdiversifier_2008} is a search-based approach
for software diversification against cyberattacks.
Given an initial instruction sequence, the algorithm generates a random
combination of the available instructions
and performs a verification test to quickly reject
non equivalent instruction sequences.
For each non-rejected sequence,
the algorithm checks semantic equivalence between the original
and the generated instruction sequences
using a SAT solver.
Superdiversifier affects the code execution time and size
by controlling the length of the generated
sequence.
A recent approach, Crow~\shortcite{arteaga2020crow}, presents a
superdiversification approach as a security mitigation for the Web.
Along the same lines,
\shortciteauthor{lundquist_searching_2016}~\citeyear{lundquist_searching_2016,lundquist2019relational}
use program synthesis
for generating program variants against
cyberattacks, but no results are available, yet.
In comparison, \ac{DivCon} uses a combinatorial
compiler backend that
measures the code quality using a more accurate
cost model that also
considers
other aspects, such as execution frequencies.

Most diversification approaches use randomized transformations in the
stack \cite{lee2021savior}, on binary code
\shortcite{wartell_binary_2012,AbrathCMBS20}, at the binary interface level
\cite{Kc03}, in the compiler \cite{homescu_large-scale_2017} or in the
source code \cite{baudry14b} to generate multiple program variants.
Unlike \ac{DivCon}, the majority of these approaches do not control
the quality of the generated variants during diversification but
rather evaluate it
afterwards~\shortcite{davi2013gadge,wang_composite_2017,koo_compiler-assisted_2018,homescu_large-scale_2017,braden_leakage-resilient_2016,crane_readactor_2015}.
However, there are a few approaches that control the code quality
during randomization.

Some compiler-based diversification approaches restrict the set of
program transformations to control the quality of the generated
code~\shortcite{crane_readactor_2015,pappas_smashing_2012}.
For example,
\shortciteauthor{pappas_smashing_2012}~\citeyear{pappas_smashing_2012}
perform software diversification at the binary level and apply three
zero-cost transformations: register randomization, instruction
schedule randomization, and function shuffling.
In contrast, \ac{DivCon}'s combinatorial approach allows it to control
the aggressiveness and potential cost of its transformations: a cost
overhead limit of 0\% forces \ac{DivCon} to apply only zero-cost
transformations; a larger limit allows \ac{DivCon} to apply more
aggressive transformations, potentially leading to higher diversity.

\shortciteauthor{homescu_profile-guided_2013}~\citeyear{homescu_profile-guided_2013}
perform only garbage (\texttt{nop}) insertion, and use a profile-guided approach
to reduce the overhead.
To do this, they control the \texttt{nop} insertion probability
based on the execution frequency of different code sections.
In contrast, \ac{DivCon}'s cost model captures different execution frequencies,
which allows it to perform more aggressive
transformations in non-critical code sections.

\section{Conclusion}
\label{sec:conclusion_fw}

This paper introduces \ac{DivCon}, a constraint-based code
diversification technique against code-reuse attacks.
The key novelty of this approach is that it supports a systematic
exploration of the trade-off between code diversity and code size and
speed.
Our experiments show that Large Neighborhood Search (\ac{LNS}) is an
effective algorithm to explore the space of diverse binary programs,
with a fine-grained control on the trade-off between code quality and
JOP gadgets diversification.
In particular, we show that the set of optimal solutions naturally
contains a set of diverse solutions, which increases significantly
when relaxing the constraint of optimality.
For improving the effectiveness of our approach against \ac{JOP}
attacks, we propose a novel gadget-specific distance measure.
Our experiments demonstrate that the diverse solutions generated by
DivCon using this distance measure are highly effective to mitigate
\ac{JOP} attacks.


\subsubsection*{Acknowledgments}
We would like to give a special acknowledgment to Christian Schulte,
for his critical contribution at the early stages of this work.
Although no longer with us, Christian continues to inspire his
students and colleagues with his lively character, enthusiasm, deep
knowledge, and understanding.
We would also like to thank Linnea Ingmar and the anonymous reviewers
of CP'2020 and JAIR for their useful feedback, and Oscar Eriksson for
proof reading.

\begin{appendices}

\section{Relax Rate Selection}
\label{sec:appa}

The \ac{LNS} configuration of \ac{DivCon} requires selecting the
\textit{relax rate}.
The relax rate is the probability that \ac{LNS} destroys
a variable at every restart,
which affects the distance between two subsequent solutions.
A higher relax rate increases diversity but requires more solving effort.
%
%

In \ac{LNS}, the relax rate, $r$, affects how many of the assigned variables of
the last solution \ac{LNS} destroys for finding the next solution.  To
evaluate that, we use two metrics and \ac{RS} as a baseline.
$P_{\delta}$ and $P_{t}$ 
correspond to the rate of the \ac{LNS} over \ac{RS} with regards to
the pairwise distance and the diversification time as follows:

\begin{equation}
  P_{\delta}(\delta, S_1, S_2) = \begin{cases}
    \displaystyle \frac{d(\delta, S_1)}{d(\delta, S_2)},& d(\delta, S_1) > d(\delta, S_2) \\
    \displaystyle \frac{d(\delta, S_2)}{d(\delta, S_1)},& otherwise
    \end{cases}
    \label{eq:impr}
\end{equation}

\noindent
and

\begin{equation}
  P_{t}(t_1, t_2) = \begin{cases}
    \displaystyle \frac{t_1}{t_2},& t_1> t_2\\
    \displaystyle \frac{t_2}{t_1},& otherwise
    \end{cases},
    \label{eq:imprt}
\end{equation}

\noindent
where $t_1$ is the diversification time for generating the solution set $S_1$ for \ac{RS}
and $t_2$ is the diversification time for generating the solution set $S_2$ for \ac{LNS}.

\begin{figure}
  \centering
  \subfloat[\label{fig:reldist}Diversity improvement]{
    \includegraphics[trim={1.5cm 0cm 2cm 0cm}, clip, width=0.5\textwidth]{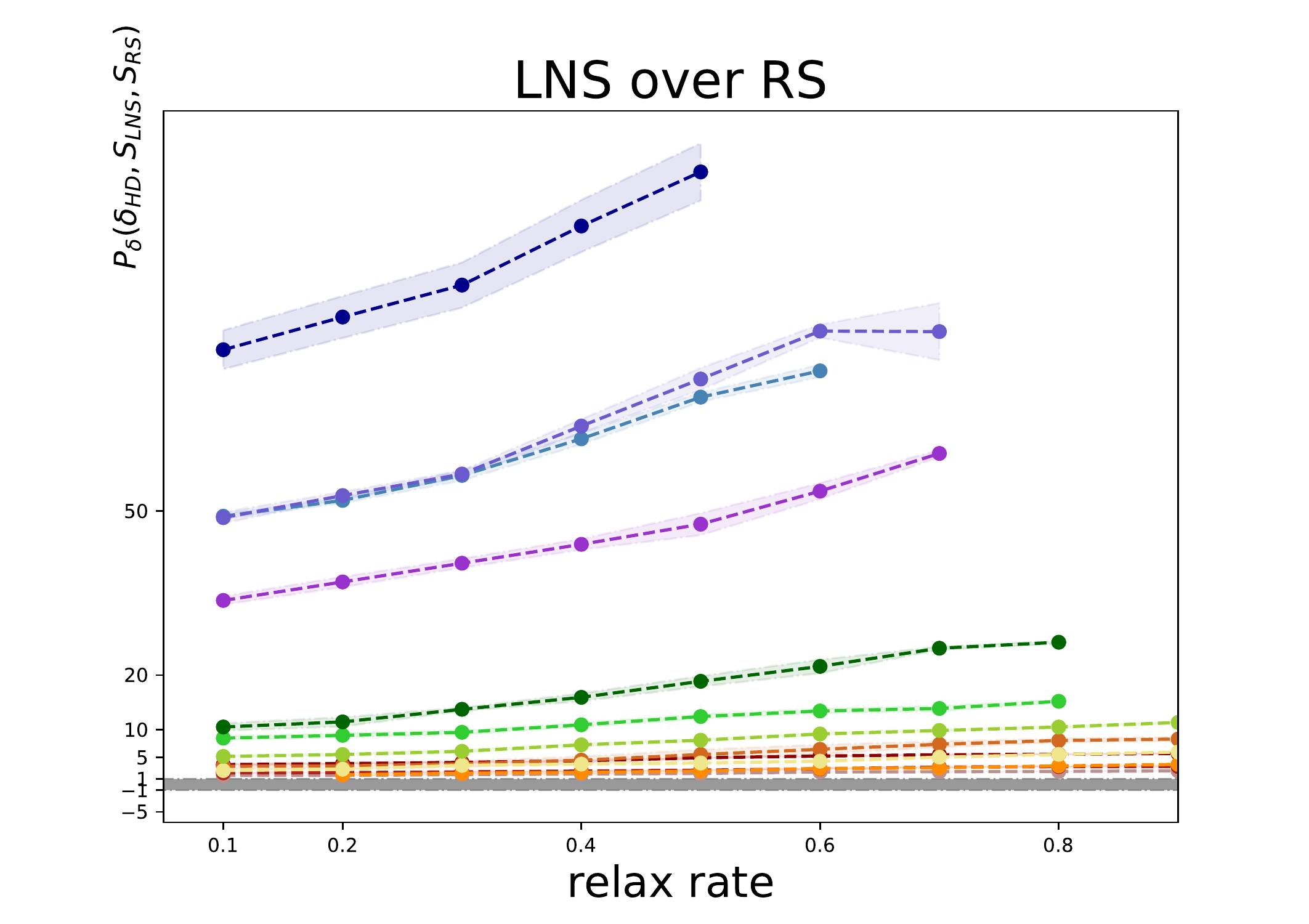}
  }%
  \subfloat[\label{fig:reltime} Diversification time overhead]{
    \includegraphics[trim={1.5cm 0cm 2cm 0cm}, clip, width=0.5\textwidth]{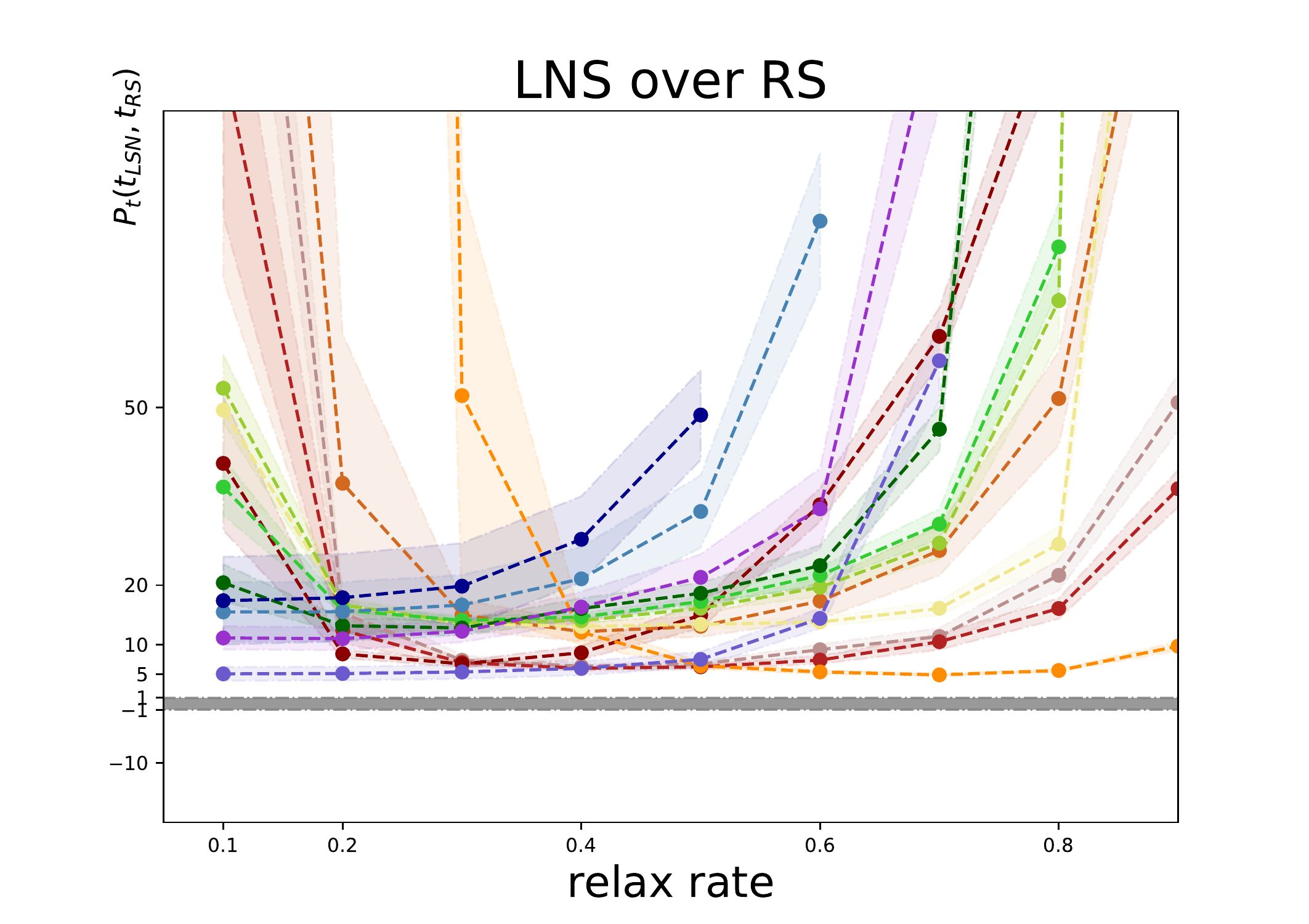}
  }%
  \caption{\label{fig:relax} Improvement in diversity and
    diversification time overhead of \ac{LNS} over \ac{RS} for
    different values of the relax rate }
\end{figure}

Figure~\ref{fig:relax} depicts the effect of different relax rates on
the distance, $P_{\delta}$, and the diversification time, $P_{t}$, when
generating 200 variants for the 14 benchmarks of
Table~\ref{tab:mediabench}.  The figure shows the results for each of
the benchmarks as a separate line with the corresponding standard deviation.
The time limit is ten minutes and the
distance measure is \acf{HD}, $\delta_{HD}$.  Figure~\ref{fig:reldist} shows that
increasing the relax rate increases the pairwise distance improvement,
$P_{\delta}$, of the generated program
variants.  Figure~\ref{fig:reltime} shows the diversification time improvement, $P_{t}$.
This figure shows that low values and large values of $r$
have large time overhead compared to \ac{RS}, whereas values $r=0.3$,
$r=0.4$, $r=0.5$, and $r=0.6$ have acceptable time overhead.  As
we have seen in Figure~\ref{fig:reldist},
the larger the relax rate, the higher the diversity improvement for
\ac{LNS} compared to \ac{RS}.  Therefore, $r=0.6$ is a good trade-off
between diversity and scalability.

\section{Diversification Example}
\label{sec:appb}

\begin{figure}[t]
  \subfloat[g721.g711.ulaw2alaw - Variant 1]{%
    \label{lst:var1}
    \begin{minipage}[t]{.48\textwidth}
\begin{lstlisting}[style=mipsstyle, firstline=13, lastline=56]
ulaw2alaw:                              # @ulaw2alaw
	.frame	$sp,0,$ra
	.mask 	0x00000000,0
	.fmask	0x00000000,0
	.set	noreorder
	.set	nomacro
	.set	noat
# BB#0:
	lui	$v0, _gp_disp
	nop
	addiu	$v0, $v0, _gp_disp
	andi	$a2, $a0, 128
	beqz	$a2, $BB0_2
	addu	$t7, $v0, $t9
# BB#1:
	!\tikzmark{copy1}!move	 $t9, $a0!\tikzmark{copy2}!
	move	 $fp, $t9
	!\tikzmark{i1}!lw	$t2, _u2a($t7)
	nop
	nop
	xori	$t5, $fp, 255!\tikzmark{i2}!
	addu	$t8, $t2, $t5
	lbu	$a1, 0($t8)
	nop
	nop
	addiu	!\tikzmark{reg11}!$t8!\tikzmark{reg12}!, $a1, -1
	b	$BB0_3
	xori	$a0, $t8, 213
$BB0_2:
	lw	$v0, _u2a($t7)
	nop
	nop
	!\tikzmark{gnb1}!xori	$a1, $a0, 127
	addu	$t0, $v0, $a1
	lbu	$t3, 0($t0)
	nop
	nop
	move	 $t6, $t3
	move	 $a0, $t6
	!\tikzmark{g3b1}!addiu	$fp, $a0, -1
	!\tikzmark{g2b1}!xori	$a0, $fp, 85
$BB0_3:
	!\tikzmark{g1b1}!jr	$ra
	andi	$v0, $a0, 255!\tikzmark{g1e1}!
\end{lstlisting}
      \begin{tikzpicture}[remember picture,overlay]
  \draw[black,thick,dotted, rounded corners]
  ([shift={(-3pt,1.5ex)}]pic cs:g1b1) 
  rectangle 
  ([shift={(3pt,-0.65ex)}]pic cs:g1e1);

  \draw[black,thick, dotted, rounded corners]
  ([shift={(-3pt,1.5ex)}]pic cs:g2b1) 
  rectangle 
  ([shift={(5pt,-0.65ex)}]pic cs:g1e1);


  \draw[black,thick, dotted, rounded corners]
  ([shift={(-3pt,1.5ex)}]pic cs:gnb1) 
  rectangle 
  ([shift={(7pt,-0.65ex)}]pic cs:g1e1);

  \draw[fill=yellow,rounded corners,opacity=0.3]
  ([shift={(-3pt,1.5ex)}]pic cs:i1) 
  rectangle 
  ([shift={(9pt,-0.65ex)}]pic cs:i2);

  \draw[fill=orange,rounded corners,opacity=0.3]
  ([shift={(-3pt,1.5ex)}]pic cs:copy1) 
  rectangle 
  ([shift={(9pt,-0.65ex)}]pic cs:copy2);

  \draw[fill=gray,rounded corners,opacity=0.3]
  ([shift={(-3pt,1.5ex)}]pic cs:reg11) 
  rectangle 
  ([shift={(9pt,-0.65ex)}]pic cs:reg12);

\end{tikzpicture}
      \vspace{-0.2cm}
    \end{minipage}
}
\subfloat[g721.g711.ulaw2alaw - Variant 2]{%
    \label{lst:var2}
    \begin{minipage}[t]{.48\textwidth}
\begin{lstlisting}[style=mipsstyle, firstline=13, lastline=56]
ulaw2alaw:                              # @ulaw2alaw
	.frame	$sp,0,$ra
	.mask 	0x00000000,0
	.fmask	0x00000000,0
	.set	noreorder
	.set	nomacro
	.set	noat
# BB#0:
	lui	$v0, _gp_disp
	nop
	addiu	$v0, $v0, _gp_disp
	andi	$fp, $a0, 128
	!\tikzmark{nop21}!addiu	$sp, $sp, 0!\tikzmark{nop22}!
	beqz	$fp, $BB0_2
	addu	$a1, $v0, $t9
# BB#1:
	!\tikzmark{i21}!xori	$t5, $a0, 255
	lw	$t6, _u2a($a1)!\tikzmark{i22}!
	nop
	nop
	addu	$v0, $t6, $t5
	lbu	$t1, 0($v0)
	nop
	nop
	addiu	!\tikzmark{reg21}!$t3!\tikzmark{reg22}!, $t1, -1
	b	$BB0_3
	xori	$fp, $t3, 213
$BB0_2:
	xori	$a3, $a0, 127
	lw	$t4, _u2a($a1)
	nop
	nop
	!\tikzmark{gnb}!addu	$gp, $t4, $a3
	lbu	$t2, 0($gp)
	nop
	nop
	addiu	$t6, $t2, -1
	!\tikzmark{g3b}!xori	$fp, $t6, 85
$BB0_3:
	!\tikzmark{g2b}!andi	$v0, $fp, 255
	!\tikzmark{g1b}!jr	$ra
	!\tikzmark{nop23}!addiu	$sp, $sp, 0!\tikzmark{g1e}!
	.set	at
	.set	macro
\end{lstlisting}
      \begin{tikzpicture}[remember picture,overlay]
  \draw[black, thick, dotted,  rounded corners]
  ([shift={(-3pt,1.5ex)}]pic cs:g1b) 
  rectangle 
  ([shift={(10pt,-0.65ex)}]pic cs:g1e);

  \draw[black, thick, dotted,  rounded corners]
  ([shift={(-3pt,1.5ex)}]pic cs:g2b) 
  rectangle 
  ([shift={(12pt,-0.65ex)}]pic cs:g1e);


  \draw[black, thick, dotted,  rounded corners]
  ([shift={(-3pt,1.5ex)}]pic cs:gnb) 
  rectangle 
  ([shift={(14pt,-0.65ex)}]pic cs:g1e);

  \draw[fill=yellow, rounded corners, opacity=0.3]
  ([shift={(-3pt,1.5ex)}]pic cs:i21) 
  rectangle 
  ([shift={(9pt,-0.65ex)}]pic cs:i22);

  \draw[fill=red, rounded corners, opacity=0.3]
  ([shift={(-3pt,1.5ex)}]pic cs:nop21) 
  rectangle 
  ([shift={(3pt,-0.65ex)}]pic cs:nop22);

  \draw[fill=red, rounded corners, opacity=0.3]
  ([shift={(-3pt,1.5ex)}]pic cs:nop23) 
  rectangle 
  ([shift={(3pt,-0.65ex)}]pic cs:g1e);

  \draw[fill=gray,rounded corners,opacity=0.3]
  ([shift={(-3pt,1.5ex)}]pic cs:reg21) 
  rectangle 
  ([shift={(9pt,-0.65ex)}]pic cs:reg22);

  \draw[<->,
    postaction={decorate,decoration=
      {text along path,
        text align=center,
        raise=1ex,
      },
      font=\small
    }]
  ([yshift=5pt]pic cs:reg12) to[bend left=10] ([yshift=5pt]pic cs:reg21) ;

  \path[draw, <->] (pic cs:i2) -> (pic cs:i21);
  \path[draw, <->] ([xshift=-110pt]pic cs:i22) -> ([xshift=110pt]pic cs:i1);

  \node[draw,
    ellipse,
    dotted,
    inner sep=0pt] (gl) at ([yshift=15pt]$(pic cs:gnb1)!0.60!(pic cs:gnb)$) {gadgets};

  \node[draw=black!30!white,,
    ellipse,
    inner sep=0pt,
    fill=gray!30!white,
    align=center,
    font=\scriptsize] (rr) at ([yshift=13pt]$(pic cs:reg12)!0.35!(pic cs:reg21)$) {register\\renaming
  };
  \node[draw=black!30!white,
    ellipse,
    inner sep=0pt,
    fill=yellow!30!white,
    align=center,
    font=\tiny] (is) at ([yshift=-16pt]$(pic cs:i2)!0.77!(pic cs:i21)$) {instruction\\reordering};

  \node[draw=black!30!white,
    ellipse,
    inner sep=2pt,
    fill=orange!30!white,
    align=center,
    font=\scriptsize] (is) at ([xshift=40pt,yshift=2pt]pic cs:copy2) {copy};

  \node[draw=black!30!white,
    ellipse,
    inner sep=2pt,
    fill=red!30!white,
    align=center,
    font=\scriptsize] (is) at ([xshift=-30pt,yshift=2pt]pic cs:nop21) {nop};

  \path[dotted, ->] (gl.-105) edge[bend left] ([xshift=5pt]pic cs:g1e1);
  \path[dotted, ->] (gl.-110) edge[bend left] ([xshift=8pt, yshift=10pt]pic cs:g1e1);
  \path[dotted, ->] (gl.-115) edge[bend left] ([xshift=8pt, yshift=50pt]pic cs:g1e1);
  \path[dotted, ->] (gl.-80) edge[bend right] ([xshift=-5pt]pic cs:g1b);
  \path[dotted, ->] (gl.-70) edge[bend right] ([xshift=-5pt, yshift=10pt]pic cs:g1b);
  \path[dotted, ->] (gl.-65) edge[bend right] ([xshift=-5pt, yshift=40pt]pic cs:g1b);
\end{tikzpicture}
      \vspace{-0.2cm}
    \end{minipage}
}
\caption{\label{fig:mips_example_ext} Example function diversification in MIPS32 assembly code}
\end{figure}
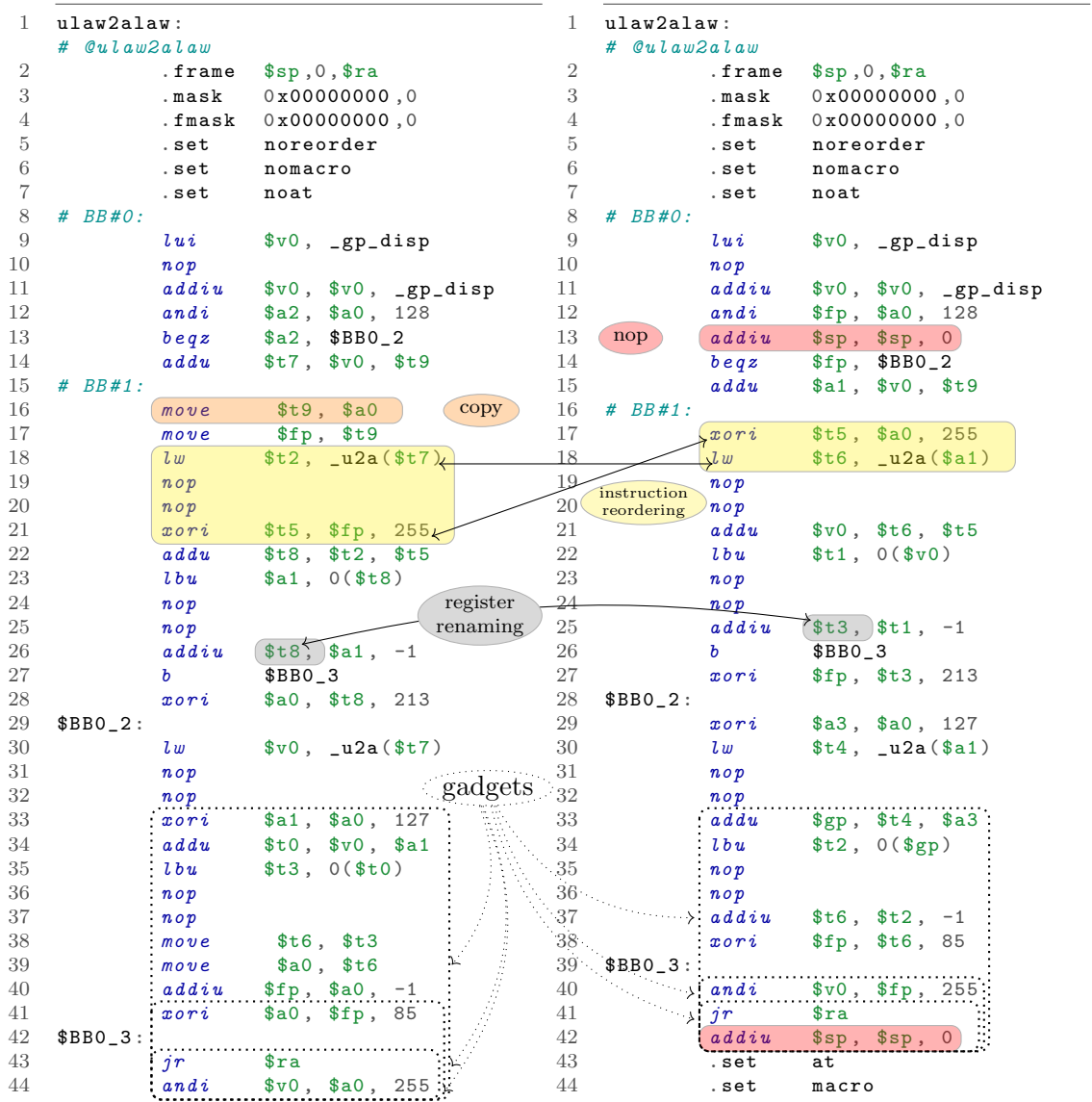

This section shows a more elaborated example of diversified code using
\ac{DivCon}.
Figure~\ref{fig:mips_example_ext} shows two variants of function
\texttt{ulaw2alaw} from application \texttt{g721}.
This function converts u-law ($\mu$-law) values to a-law ($A$-law) values.
$\mu$-law and $A$-law are the two main companding algorithms of
G.711~\cite{itu729general}.

The two variants, Listing~\ref{lst:var1} and Listing~\ref{lst:var2},
are generated by \ac{DivCon} with relax rate 0.6,  optimality gap 10\%,
and the cycle hamming distance, $\delta_{HD}$.
Figure~\ref{fig:mips_example_ext} highlights four different ways in
which the two variants differ.
First, \ac{DivCon} may add \texttt{nop} instructions.
Interestingly, \ac{DivCon} added an empty stack frame to Variant 2.
%
%
The prologue (line 13 in Variant 2) and epilogue (line 42 in of Variant 2)
instructions that build and destroy the empty stack frame are no-operations,
however they contribute to the diversification of the function and their
overhead does not exceed the optimality gap.
Otherwise, \ac{DivCon} adds MIPS \texttt{nop} instructions to fill the
instruction schedule empty slots including the instruction delays due
to their execution latency (see lines 19 and 20 of Variant 1).
Another transformation is the addition of \texttt{copy} operations to
move data from one register to the other (highlighted at line 16 of
Variant 1).
This transformation assists register renaming, which improves
diversification.
The third transformation that we have highlighted (lines 18-21 of
Variant 1 and lines 17-18 of Variant 2) is instruction reordering.
Here, whenever there is no data dependency between the instructions,
the order of the instructions might change.
Finally, the register assignment of different operations differs, with
an example highlighted at line 26 of Variant 1 and line 25 of Variant
2.
Other transformations, like spilling to the stack, are also possible.
The function of Figure~\ref{fig:mips_example_ext} is small and does
not require spilling.
However, \ac{DivCon} may enable spilling if the overhead is not more
than the allowed optimality gap (10\% here).

Figure~\ref{fig:mips_example_ext} shows also some of the gadgets
that are available in function \texttt{ulaw2alaw}.
In particular, both variants contain a number of gadgets that all
include the last gadget, consisting of a return jump, \texttt{jr}, and
its delay slot (following the branch).
No pair of gadgets in the two variants is identical with
regards to the content or the position in the code.


\end{appendices}

\vskip 0.2in
\bibliography{bibliography}
\bibliographystyle{theapa}

\end{document}